\documentclass[structabstract]{aa}
\usepackage{graphicx}
\usepackage{lscape}
\usepackage{amsmath}
\usepackage{txfonts}
\usepackage{url}
\usepackage{natbib}
\usepackage{xcolor}

\graphicspath{{./}}
\begin{document}
\title{Populations of rotating stars}
\subtitle{II. Rapid rotators and their link to Be-type stars} 
\author{A.~Granada\inst{1}, S.~Ekstr\"om\inst{1}, C.~Georgy\inst{2,3}, J.~Krti\v cka\inst{4}, S.~Owocki\inst{5}, G.~Meynet\inst{1}, A.~Maeder\inst{1}}
\authorrunning{Granada et al.}
\institute{Geneva Observatory, University of Geneva, Maillettes 51, CH-1290 Sauverny, Switzerland\\
\email{anahi.granada@unige.ch}\\
\and
Astrophysics group, EPSAM, Keele University, Lennard-Jones Labs, Keele, ST5 5BG, UK\\
\and
Centre de Recherche Astrophysique, Ecole Normale Sup\'erieure de Lyon, 46, All\'ee d'Italie, F-69384 Lyon cedex 07, France\\
\and 
Masaryk University, Kotl\'a{\v r}sk\'a 2, CZ-611 37 Brno, Czech Republic\\
\and 
Bartol Research Institute, University of Delaware, Newark, DE 19716, USA\\}
\date{Received ; accepted } 
\abstract
   {Even though it is broadly accepted that single Be stars are rapidly rotating stars surrounded by a flat rotating circumstellar disk,
 there is still a debate about how fast these stars rotate and also about the mechanisms involved in the  angular-momentum and mass input in the disk.}
 {We study the properties of stars that rotate near their critical-rotation rate and investigate the properties of the disks formed by equatorial mass ejections.}
  {We used the most recent Geneva  stellar evolutionary tracks for rapidly rotating stars that reach the critical limit and used a simple model for the disk structure.}
 {We obtain that for a 9 $M_{\sun}$ star at solar metallicity, the minimum average velocity during the Main Sequence phase to reach the critical velocity is around 330 km s$^{-1}$, 
whereas it would be 390 km s$^{-1}$ at the metallicity of the Small Magellanic Cloud (SMC). 
 Red giants or supergiants originating from very rapid rotators rotate six times faster and show N/C ratios three times higher than those originating from slowly rotating stars.
 This difference becomes stronger at lower metallicity. It might therefore be very interesting to study the red giants in clusters that show a large number of Be stars on the MS band.
 On the basis of our single-star models, we show that the observed Be-star fraction with cluster age is compatible with the existence of a temperature-dependent lower limit in the velocity rate required for a star to become a Be star. 
 The mass, extension, and diffusion time of the disks produced when the star is losing mass at the critical velocity, obtained from simple parametrized expressions,
 are estimated to be between $9.4\cdot10^{-12}$ and $1.4\cdot10^{-7}\,M_{\sun}$ ($3\cdot10^{-6}$ to $4.7\cdot10^{-2}$ times the mass of the Earth), 2000 and 6500 $R_{\sun}$, and 10 and 30 yr. These values are not too far from those estimated for disks around Be-type stars.  At a given metallicity, the mass and the extension of the disk
increase with the initial mass and with age on the MS phase. Denser disks are expected in low-metallicity regions.
} 
 {}

\keywords{stars: general -- stars: evolution -- stars: rotation -- stars: Be -- stars: mass-loss}

\maketitle
\section{Introduction}

There is general consensus that classical Be stars are intrinsically rapidly rotating B-type stars, with analogs among O and early- A types \citep[although they are much less frequent, see {\it e.g.}][]{Zorec2005a}, surrounded by  a gaseous decretion viscous disk. Such a disk, geometrically thin and in nearly Keplerian rotation, can explain many of the characteristics that define Be stars \citep[see for example the reviews by \citealt{Porter2003a} and][]{Carciofi2011a}.

However, there are still many open questions regarding Be stars: do they reach the critical limit, or do they rotate at 70-80\% of their critical limit?  Does binarity play a role in the appearance of the Be phenomenon? Which are the mechanisms involved in the angular-momentum and mass input in the disk? As noted by \citet{Stee2009a}, it seems that even if we consider the single stars that present the Be phenomenon, they do not all share a common mass-ejection mechanism.

From an observational point of view, some authors \citep{Fremat2005a,Cranmer2005a,Cranmer2009a,Huang2010a} found that Be stars, in particular early-type Be stars, seem to be rotating below their critical limit, though in some cases the determination of rotational velocities could be underestimated \citep{Townsend2004a}. 
Also, \citet{Rivinius2001a} found for some early-type Be stars such as $\mu$ Cen that the beating of multi-periodic non-radial pulsations seems to be linked with the mass outbursts. 
Nevertheless, the lack of observational evidence for large-amplitude pulsations or other {\it strong} mechanisms that could help to drive angular momentum and mass into a circumstellar disk in many of the stars, favours the idea of weaker processes that would be able to overcome the effective gravity and let the matter escape from the photosphere into the disk: thus rotation close to the critical limit is expected \citep[\textit{e.g.}][]{Howarth2007a}. This idea is also supported by many interferometric observations, which have allowed to determine the oblateness of some Be stars \citep{Domiciano2003a,Zhao2011a,Touhami2011a,Meilland2012a,Kraus2012a}, suggesting that some of them are close to their critical rotation. 

Interestingly, theoretical studies on stellar evolution show that along the main-sequence (MS) evolution, the single intermediate-mass stars with a sufficiently high rotational velocity on the zero-age main sequence (ZAMS) can reach equatorial velocities near the critical value\footnote{By critical velocity we mean here the velocity required for the centrifugal force to counterbalance the gravity at the equator of the star \citep[see {\it e.g.}][]{Maeder2000a}.}, thanks to the transfer of angular momentum from the inner contracting parts to the outer regions \citep{Meynet2007a,Ekstrom2008b}. 

\citet{Ekstrom2008b} studied how stars with different masses, different metallicities, and different initial rotational velocities evolve towards the critical limit. In that work, the computation stopped when the model reached the critical velocity. In the study we present here, the following changes and improvements with respect to \citet{Ekstrom2008b} were made:
\begin{itemize}
\item The numerical method for computing the surface velocity as resulting from the various transport processes and mass loss was improved \citep[as explained in detail in][hereafter Paper I]{Georgy2012c}. In contrast to the computation by \citet{Ekstrom2008b}, in which the deformation of the star and the angular momentum contained in the envelope\footnote{The envelope is the region where convection is treated in a non-adiabatic way and ionisation may not be complete.} were neglected, we account here for both effects. The envelope contains one ten-thousandth of the total stellar mass during the MS phase. We assumed that this region has the same angular velocity as the first layer of the interior\footnote{The stellar interior is defined as the region where convection is adiabatic and ionisation is complete.}. This development allows for a much better tracing of the evolution of the total angular-momentum content of the star. As a result, our models now reach the critical limit less easily than in \citet{Ekstrom2008b}, as discussed in Sect.~\ref{secprop}. In addition, the effects of anisotropic winds were accounted for in the computations following \citet{Georgy2011a}.
\item The computations were now pursued all along the MS phase until the helium flash for stars with $M_\text{ini}\le 2\,M_{\sun}$, the early asymptotic giant branch for $2.5\,M_{\sun}\le M_\text{ini}\lesssim 9\,M_{\sun}$ or the end of core-C burning when $M_\text{ini} \gtrsim 9\, M_{\sun}$, even for the models that reach the critical limit. This allows us to explore the consequences of fast rotation beyond the MS phase.
\item In the present work, the physical ingredients are the same as in the paper by \citet{Ekstrom2012a}, and differ from those used by \citet{Ekstrom2008b}, as discussed in Sect.~\ref{secprop}.
\end{itemize}

Using this new version of the Geneva stellar evolution code, our group has produced grids of stellar-evolution tracks at solar metallicity ($Z=0.014$) in the mass range of 0.8 to 120 $M_{\sun}$, both rotating and non-rotating \citep{Ekstrom2012a}, and showed that the models rotating with $V_\text{ini}/V_\text{crit}$ =0.4 provide a reasonable fit to important observed characteristics of stars. 
In \defcitealias{Georgy2012c}{Paper I}\citetalias{Georgy2012c} we presented a grid to explore in detail the mass domain between $1.7$ to $15\,M_{\sun}$, which corresponds to early-A, -B, and late-O  spectral types, at different metallicities and with initial angular velocity between $0<\Omega_\text{ini}/\Omega_\text{crit}<0.95$. In this last paper, the effects of rotation on the tracks in the HR diagram, the lifetimes, the surface velocities, and abundances were discussed.

Here we specifically focus on a sub-set of these models: those whose surface velocity at the equator reaches the critical limit during the MS phase. With the study of these models, we aim to address the following questions: what can be said about the mechanical mass-loss rates when the star reaches the critical limit? What would be the extension and lifetime of a stationary circumstellar viscous decretion disk formed by matter expelled from the star? What is the mass-loss rate expected from the star-disk system? How do the properties of the disk vary as a function of the mass, metallicity and age of the star? To find the answer, we used the analytical development in \citet{Krticka2011a} together with the results presented in \citetalias{Georgy2012c}.

The paper is organised as follows: the main physical ingredients of the stellar models are briefly recalled in Sect.~\ref{secmod}. In Sect.~\ref{secprop}, we discuss the properties of models that reach the critical velocity during their MS phase: average values for the rotation velocity, properties of red giants and supergiants originating from very rapid rotators, and the frequency of rapid rotators in clusters of various ages. We discuss the mechanical mass-loss rates obtained when the models reach the critical limit in Sect.~\ref{secmec}. Section~\ref{secdisks} discusses the properties of the disks around models that rotate at the critical limit using the disk model described in \citet{Krticka2011a}. The possible relations between stars at the critical limit and classical Be-type stars are discussed in Sect.~\ref{secbe}. Finally, the main conclusions are presented in Sect.~\ref{secconclu}.

\section{Stellar models \label{secmod}}

The stellar models analysed here have masses between $1.7$ and $15\ M_{\sun}$, and metallicities of $Z=0.002$ ($Z_\text{SMC}$), $0.006$ ($Z_\text{LMC}$), and $0.014$ ($Z_{\sun}$). They are a sub-set of the grid presented in \citetalias{Georgy2012c}, since we consider here only the models that reach the critical limit during the MS phase, {\it i.e.} models with $\omega_\text{ini}=\Omega_\text{ini}/\Omega_\text{crit}\gtrsim0.80$. A detailed description of the physics included in the models can be found in \citetalias{Georgy2012c}. We briefly recall a few relevant inputs:

\begin{itemize}
\item {\bf Overshooting:} the convective-core baundaries are determined with the Schwarzschild criterion. The convective core is extended with an overshoot parameter $d_\text{over}/H_P=0.10$ \citep[after calibration, as explained in][]{Ekstrom2012a}.
\item {\bf Radiative mass loss:} the more massive models ($M>5\,M_{\sun}$) lose mass through line-driven winds, according to the mass-loss prescription by \citet{deJager1988a}. Only stars with an initial mass greater than 12 $M_{\sun}$ 
lose a non-negligible amount of mass through radiative winds during the MS phase. For instance, our 15$\,M_{\sun}$ stellar model at $Z=0.014$ loses about 10\% of its initial mass during the MS phase through radiative winds. 
For the red-giant phase, the mass loss rates are obtained by the prescription of \citet{Reimers1975,Reimers1977} up to 12$\,M_{\sun}$, whereas the \citet{deJager1988a} prescription is used for masses of 15 $\,M_{\sun}$ or larger.
\item {\bf Mechanical mass loss:} when overcritical velocities are reached at the equator of a star, some matter is released and presumably launched along Keplerian orbits in a disk around the star. In the following, we call such an event an overcritical episode\footnote{It is important to properly distinguish between $V/V_\text{crit}$ and $\Omega/\Omega_\text{crit}$. Both quantities indicate how close to its breakup limit a star rotates, but they are \textit{not} equivalent. Writing the angular velocity as $\Omega\,=\,V_\text{eq}/R_\text{eq}$, where $V_\text{eq}$ is the rotational velocity at the equator and $R_\text{eq}$ is the equatorial radius of the star, it turns out that $\Omega/\Omega_\text{crit}\,=\,(V_\text{eq}/V_\text{crit})\cdot(R_\text{eq,crit}/R_\text{eq})$. Because $R_\text{eq}$ depends on the rotational velocity of the star, $\Omega/\Omega_\text{crit}$ is equal to $V_\text{eq}/V_\text{crit}$ only when  $\Omega=0$ or in a star rotating at the critical limit $R_\text{eq,crit}\,=\,R_\text{eq}$. Throughout this work, every time we refer to rotation close to the critical limit, we implicitly assume that we talk about $\Omega/\Omega_\text{crit}$.}. Because of numerical difficulties in computing of the stellar evolution when the model is at the critical limit, we need to define a value for the highest fraction of the critical angular velocity allowed, $\omega_\text{max}$, above which the mechanical mass loss is switched on. In our computation, we set $\omega_\text{max}=0.99$. The quantity of mass ejected mechanically from the model is governed by the quantity of angular momentum that needs to be removed from the surface to maintain the model below the limit set by $\omega_\text{max}$, as explained in \citetalias{Georgy2012c}. The mass injection into the disk allows the star to decrease its surface velocity below the critical value, until the secular evolution (and more precisely the transport of angular momentum by meridional circulation and shears) drives the surface to become overcritical again. Thus, once the critical rotation is reached, the surface velocity of the model will stay around the critical value until the end of the MS, {\it i.e.} until the evolutionary timescale becomes too short to allow the transport of angular momentum to the surface.
\item {\bf Rotation:} the transport of angular momentum and chemical species inside a star is implemented following the prescription of \citet{Zahn1992a} for the horizontal diffusion coefficient, and that of  \citet{Maeder1997a} for the shear diffusion coefficient.
\end{itemize}

\section{Properties of stars at the critical limit \label{secprop}}

\begin{figure}
\centering{
\includegraphics[width=.5\textwidth]{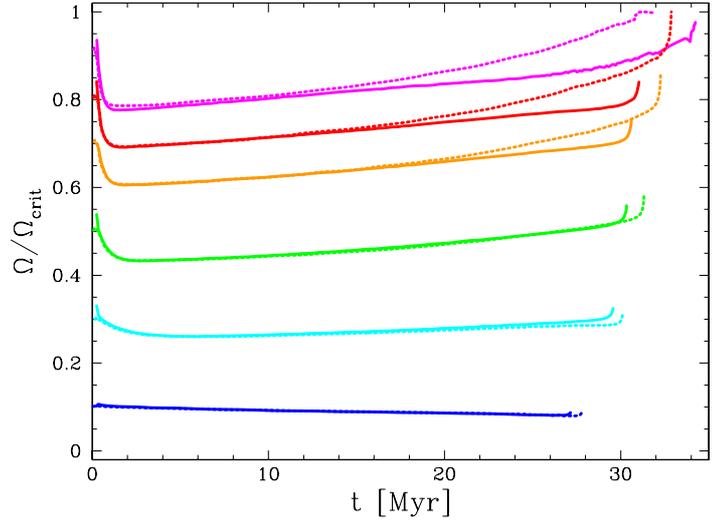}}
\caption{Evolution of the $\Omega/\Omega_\text{crit}$ ratio for 9 $M_{\sun}$ at $Z=0.002$, with initial values of $\Omega_\text{ini}/\Omega_\text{crit}=$ 0.10, 0.30, 0.50, 0.70, 0.80, and 0.90 (bottom to top) during the MS phase. The solid lines show the present models, the dotted lines the models by \citet{Ekstrom2008b}. }
\label{comp9}
\end{figure}

In Fig.~\ref{comp9}, we show the evolution of the $\omega=\Omega/\Omega_\text{crit}$ ratio for the 9 $M_{\sun}$ models at the SMC metallicity, together with the corresponding models of \citet[dotted lines]{Ekstrom2008b}. 

We recall that we start the computation of a model with solid-body rotation on the ZAMS, similarly to the work by \citet{Heger2000a}. The underlying assumption is that pre-MS stars are fully convective, and that convection drives rigid-body rotation. Once the solid-body constraint is released, the flat internal profile of rotation is modified very rapidly under the action of meridional currents, and the surface rotation decreases strongly, as can be clearly seen in Fig.~\ref{comp9}. This means that, in the present scenario, the models begin their evolution with a slow-down of their surface rotational velocity, and may reach the critical limit only after some evolution on the MS. As a result, the present way of initiating the computation makes it impossible to obtain very rapidly rotating stars at a very young age, even when imposing a high value of $\omega_\text{ini}$. The validity of the assumption made is now being explored with models of pre-MS evolution and will be the subject of a future publication. The first results show that the internal rotation profile depends to some extent on the way the star mass and angular-momentum content build up through accretion. At the moment, though, we take the present results at face value, keeping in mind that we might be missing some young critical rotators.

From Fig.~\ref{comp9}, we also see that the critical velocity is reached at the end of the MS phase. This occured in the models by \citet{Ekstrom2008b} and is confirmed by the present models, although
significant differences between these two sets of models appear for the most rapidly rotating models. The change in the prescription for the diffusion coefficient explains the slight changes in the MS lifetimes, but has no impact on the evolution of the surface velocity. This is clearly shown in the work by \citet{Meynet2013}, who 
compared the time-averaged velocities during the MS phase obtained using different sets of diffusion coefficients, D$_{shear}$ and D$_{h}$. These authors found that for a 15 $M_{\sun}$ model at $Z=0.002$ with V$_{ini}$/V$_{crit}$=0.5 the differences caused by using the present prescription and that of \citet{Ekstrom2008b} are very small, of the order of only 0.5\%.
 The source of the difference in surface velocities for the most rapidly rotating models is the improved numerical procedure for computing the surface velocity. The present models maintain a lower surface velocity because we now account for the angular momentum in the envelope and for the deformation of the star, and we also ensure the angular-momentum conservation of the star+wind system (including radiative and mechanical mass loss). As a result, the present models evolve less easily towards the critical limit, while the models by \citet{Ekstrom2008b} overestimated the surface velocity.

The main properties of the models reaching the critical velocity at their surface during the MS phase are summarised in Table~\ref{TabBeModels}. 

From Table~\ref{TabBeModels}, we see that only models with $\omega_\text{ini}>0.80-0.90$ are able to reach the critical velocity during the MS phase. If we compare these limits with those given by \citet{Ekstrom2008b}, we see that for the metallicity and the mass range in common between the two works, namely for the 3 and 9 $M_{\sun}$ at $Z=0.002$, the present models require a minimum $\omega_\text{ini}=0.95$ , while the models by \citet{Ekstrom2008b} require $\omega_\text{ini}=0.90$ and $0.80$, respectively. This reflects the fact mentioned above that the present models approach the critical limit less easily.
We also see that the lower-mass models reach the critical rotation much more easily than the more massive ones and can spend up to $30\%$ of the total MS time at the critical velocity. This trend was also obtained in \citet{Ekstrom2008b} at $Z=0.002$. This comes from the longer time available in lower-mass models, for meridional circulation and shears to bring angular momentum at the surface. Moreover, in the higher initial-mass models, mass loss by stellar winds may remove some angular momentum from the surface and thus keep the star away from the critical limit (although this effect is quite weak at low metallicity).

\begin{table*}
\caption{Data for the models that reach the critical velocity during the MS.}
\centering
\begin{tabular}{crc||ccccccccc}
\hline\hline
\rule[0mm]{0mm}{3mm}$Z$ & $M_\text{ini}$ & $\Omega/\Omega_\text{crit, ini}$ & $\tau_\text{crit. rot.}$ & $\tau_\text{crit. rot.}/\tau_\text{MS}$ & $H_\text{cen}$ & $\Delta M_\text{tot}$ & $\Delta M_\text{rad}$ & $\Delta M_\text{mech}$ & $\dot{M}_\text{mech, mean}$ &$\overline{V}_\text{eq}$ & f(N/C) \\
\rule[-1.5mm]{0mm}{2mm} & $M_{\sun}$ & & $\text{yr}$ & & & $M_{\sun}$ & $M_{\sun}$ & $M_{\sun}$ & $M_{\sun}\cdot \text{yr}^{-1}$ &  km s$^{-1}$ &  \\
\hline
\rule[0mm]{0mm}{3mm}$0.014$ & $  1.70$ & $0.90$ & $1.5123\cdot 10^{7}$ & $0.007$ & $0.015$ & $0.1006$ & $0.1002$ & $3.8893\cdot 10^{-4}$ & $2.5717\cdot 10^{-11}$ & 243& 2.0\\
    $0.014$ & $  1.70$ & $0.95$ & $6.3372\cdot 10^{8}$ & $0.290$ & $0.325$ & $0.1008$ & $0.0990$ & $1.8402\cdot 10^{-3}$ & $2.9038\cdot 10^{-12}$ & 265& 2.5\\
    $0.014$ & $  2.00$ & $0.90$ & $6.0923\cdot 10^{7}$ & $0.046$ & $0.085$ & $0.0456$ & $0.0445$ & $1.1583\cdot 10^{-3}$ & $1.9012\cdot 10^{-11}$ & 253& 2.4\\
    $0.014$ & $  2.00$ & $0.95$ & $2.4103\cdot 10^{8}$ & $0.180$ & $0.240$ & $0.0473$ & $0.0454$ & $1.9232\cdot 10^{-3}$ & $7.9793\cdot 10^{-12}$ & 263& 2.7\\
    $0.014$ & $  2.50$ & $0.90$ & $1.6892\cdot 10^{7}$ & $0.024$ & $0.053$ & $0.0530$ & $0.0519$ & $1.0888\cdot 10^{-3}$ & $6.4456\cdot 10^{-11}$ & 257& 2.6\\
    $0.014$ & $  2.50$ & $0.95$ & $1.4935\cdot 10^{8}$ & $0.212$ & $0.273$ & $0.0533$ & $0.0500$ & $3.2941\cdot 10^{-3}$ & $2.2057\cdot 10^{-11}$ & 281& 3.5\\
    $0.014$ & $  3.00$ & $0.90$ & $1.3361\cdot 10^{7}$ & $0.032$ & $0.068$ & $0.1106$ & $0.1093$ & $1.2727\cdot 10^{-3}$ & $9.5256\cdot 10^{-11}$ & 267& 3.0\\
    $0.014$ & $  3.00$ & $0.95$ & $9.2771\cdot 10^{7}$ & $0.220$ & $0.283$ & $0.1356$ & $0.1315$ & $4.0566\cdot 10^{-3}$ & $4.3727\cdot 10^{-11}$ & 293& 4.0\\
    $0.014$ & $  4.00$ & $0.90$ & $3.9870\cdot 10^{6}$ & $0.020$ & $0.046$ & $0.1032$ & $0.1013$ & $1.9081\cdot 10^{-3}$ & $4.7859\cdot 10^{-10}$ & 284& 4.0\\
    $0.014$ & $  4.00$ & $0.95$ & $3.3829\cdot 10^{7}$ & $0.171$ & $0.241$ & $0.1093$ & $0.1035$ & $5.8012\cdot 10^{-3}$ & $1.7149\cdot 10^{-10}$ & 312& 5.5\\
    $0.014$ & $  5.00$ & $0.90$ & $1.4965\cdot 10^{6}$ & $0.013$ & $0.031$ & $0.1019$ & $0.1001$ & $1.7413\cdot 10^{-3}$ & $1.1636\cdot 10^{-9}$  & 298& 5.4\\
    $0.014$ & $  5.00$ & $0.95$ & $1.6155\cdot 10^{7}$ & $0.142$ & $0.208$ & $0.1097$ & $0.1032$ & $6.4987\cdot 10^{-3}$ & $4.0227\cdot 10^{-10}$ & 326& 7.4\\
    $0.014$ & $  7.00$ & $0.90$ & $4.8294\cdot 10^{5}$ & $0.009$ & $0.019$ & $0.1337$ & $0.1316$ & $2.1063\cdot 10^{-3}$ & $4.3614\cdot 10^{-9}$ & 320& 11.5\\
    $0.014$ & $  7.00$ & $0.95$ & $6.0779\cdot 10^{6}$ & $0.115$ & $0.167$ & $0.1435$ & $0.1352$ & $8.2297\cdot 10^{-3}$ & $1.3540\cdot 10^{-9}$ & 348& 15.2\\
    $0.014$ & $  9.00$ & $0.90$ & $1.5793\cdot 10^{5}$ & $0.005$ & $0.010$ & $0.3860$ & $0.3851$ & $8.9614\cdot 10^{-4}$ & $5.6743\cdot 10^{-9}$ & 335& 19.0\\
    $0.014$ & $  9.00$ & $0.95$ & $2.2474\cdot 10^{6}$ & $0.069$ & $0.109$ & $0.3134$ & $0.3039$ & $9.4472\cdot 10^{-3}$ & $4.2036\cdot 10^{-9}$ & 364& 24.4\\
\hline
 \rule[0mm]{0mm}{3mm}$0.006$ & $  1.70$ & $0.80$ & $2.3191\cdot 10^{6}$ & $0.001$ & $0.001$ & $0.0853$ & $0.0853$ & $7.4060\cdot 10^{-6}$ & $3.1934\cdot 10^{-12}$ & 219& 2.7\\
    $0.006$ & $  1.70$ & $0.90$ & $1.5500\cdot 10^{8}$ & $0.092$ & $0.148$ & $0.0882$ & $0.0867$ & $1.4488\cdot 10^{-3}$ & $9.3468\cdot 10^{-12}$ & 262& 4.4\\
    $0.006$ & $  1.70$ & $0.95$ & $5.0047\cdot 10^{8}$ & $0.295$ & $0.339$ & $0.0879$ & $0.0845$ & $3.4563\cdot 10^{-3}$ & $6.9061\cdot 10^{-12}$ & 284& 5.5\\
    $0.006$ & $  2.00$ & $0.80$ & $9.0662\cdot 10^{4}$ & $0.000$ & $0.000$ & $0.0321$ & $0.0321$ & $1.0098\cdot 10^{-5}$ & $1.1138\cdot 10^{-10}$ & 224& 3.2\\
    $0.006$ & $  2.00$ & $0.90$ & $7.2708\cdot 10^{7}$ & $0.069$ & $0.124$ & $0.0367$ & $0.0353$ & $1.4306\cdot 10^{-3}$ & $1.9676\cdot 10^{-11}$ & 269& 5.0\\
    $0.006$ & $  2.00$ & $0.95$ & $2.6859\cdot 10^{8}$ & $0.255$ & $0.314$ & $0.0368$ & $0.0332$ & $3.6382\cdot 10^{-3}$ & $1.3546\cdot 10^{-11}$ & 292& 6.5\\
    $0.006$ & $  2.50$ & $0.90$ & $8.5848\cdot 10^{6}$ & $0.015$ & $0.034$ & $0.0537$ & $0.0528$ & $8.4155\cdot 10^{-4}$ & $9.8028\cdot 10^{-11}$ & 271& 5.1\\
    $0.006$ & $  2.50$ & $0.95$ & $1.0671\cdot 10^{8}$ & $0.189$ & $0.260$ & $0.0458$ & $0.0426$ & $3.1821\cdot 10^{-3}$ & $2.9820\cdot 10^{-11}$ & 298& 7.3\\
    $0.006$ & $  3.00$ & $0.90$ & $3.5621\cdot 10^{6}$ & $0.010$ & $0.021$ & $0.0949$ & $0.0943$ & $6.4225\cdot 10^{-4}$ & $1.8030\cdot 10^{-10}$ & 281& 5.8\\
    $0.006$ & $  3.00$ & $0.95$ & $4.1819\cdot 10^{7}$ & $0.120$ & $0.191$ & $0.1043$ & $0.1013$ & $2.9769\cdot 10^{-3}$ & $7.1187\cdot 10^{-11}$ & 307& 8.4\\
    $0.006$ & $  4.00$ & $0.90$ & $4.6283\cdot 10^{5}$ & $0.003$ & $0.006$ & $0.0914$ & $0.09134$ & $5.7245\cdot 10^{-5}$ & $1.2369\cdot 10^{-10}$ &295 & 8.4\\
    $0.006$ & $  4.00$ & $0.95$ & $4.9446\cdot 10^{6}$ & $0.029$ & $0.062$ & $0.1076$ & $0.1052$ & $2.3220\cdot 10^{-3}$ & $4.6961\cdot 10^{-10}$ & 322& 11.4\\
    $0.006$ & $  5.00$ & $0.90$ & $1.4823\cdot 10^{5}$ & $0.001$ & $0.001$ & $0.0988$ & $0.0988$ & $1.0418\cdot 10^{-5}$ & $7.0283\cdot 10^{-11}$ & 307& 14.7\\
    $0.006$ & $  5.00$ & $0.95$ & $8.0123\cdot 10^{5}$ & $0.008$ & $0.017$ & $0.0958$ & $0.0944$ & $1.4447\cdot 10^{-3}$ & $1.8031\cdot 10^{-9}$ & 334& 16.1\\
    $0.006$ & $  7.00$ & $0.95$ & $2.1137\cdot 10^{5}$ & $0.004$ & $0.008$ & $0.0489$ & $0.0479$ & $1.0464\cdot 10^{-3}$ & $4.9506\cdot 10^{-9}$ &355 & 41.8\\
    $0.006$ & $  9.00$ & $0.95$ & $8.4138\cdot 10^{5}$ & $0.025$ & $0.052$ & $0.1450$ & $0.1401$ & $4.8920\cdot 10^{-3}$ & $5.8142\cdot 10^{-9}$ & 372& 116.8\\
    $0.006$ & $ 12.00$ & $0.90$ & $1.5667\cdot 10^{5}$ & $0.008$ & $0.015$ & $1.2625$ & $1.2596$ & $2.9459\cdot 10^{-3}$ & $1.8803\cdot 10^{-8}$ & 360& 102.7\\
    $0.006$ & $ 12.00$ & $0.95$ & $3.0515\cdot 10^{5}$ & $0.015$ & $0.048$ & $0.8632$ & $0.8557$ & $7.4742\cdot 10^{-3}$ & $2.4494\cdot 10^{-8}$ & 391& 155.7\\
\hline
 \rule[0mm]{0mm}{3mm}$0.002$ & $  1.70$ & $0.90$ & $3.3156\cdot 10^{7}$ & $0.023$ & $0.006$ & $0.0735$ & $0.0735$ & $1.5584\cdot 10^{-5}$ & $4.7001\cdot 10^{-13}$ & 267& 18.2\\
    $0.002$ & $  1.70$ & $0.95$ & $2.5469\cdot 10^{8}$ & $0.150$ & $0.283$ & $0.0559$ & $0.0534$ & $2.4547\cdot 10^{-3}$ & $9.6380\cdot 10^{-12}$ & 298& 44.1\\
    $0.002$ & $  2.00$ & $0.90$ & $6.5993\cdot 10^{6}$ & $0.008$ & $0.017$ & $0.0335$ & $0.0333$ & $2.1948\cdot 10^{-4}$ & $3.3258\cdot 10^{-11}$ & 282& 13.0\\
    $0.002$ & $  2.00$ & $0.95$ & $9.9121\cdot 10^{7}$ & $0.114$ & $0.166$ & $0.0346$ & $0.0327$ & $1.8645\cdot 10^{-3}$ & $1.8810\cdot 10^{-11}$ & 308& 20.0\\
    $0.002$ & $  2.50$ & $0.90$ & $7.4248\cdot 10^{5}$ & $0.002$ & $0.000$ & $0.0733$ & $0.0733$ & $1.2505\cdot 10^{-5}$ & $1.6842\cdot 10^{-11}$ & 285& 17.7\\
    $0.002$ & $  2.50$ & $0.95$ & $9.5637\cdot 10^{6}$ & $0.020$ & $0.023$ & $0.0883$ & $0.0879$ & $4.6619\cdot 10^{-4}$ & $4.8746\cdot 10^{-11}$ & 310& 20.0\\
    $0.002$ & $  3.00$ & $0.95$ & $8.9608\cdot 10^{5}$ & $0.003$ & $0.005$ & $0.0860$ & $0.0859$ & $7.6874\cdot 10^{-5}$ & $8.5789\cdot 10^{-11}$ & 319& 26.5\\
    $0.002$ & $  7.00$ & $0.95$ & $1.5529\cdot 10^{5}$ & $0.003$ & $0.006$ & $0.0375$ & $0.3705$ & $4.5940\cdot 10^{-4}$ & $2.9584\cdot 10^{-9}$ & 365& 325.0\\
    $0.002$ & $  9.00$ & $0.95$ & $4.5056\cdot 10^{5}$ & $0.013$ & $0.042$ & $0.1100$ & $0.1025$ & $7.5032\cdot 10^{-3}$ & $1.6653\cdot 10^{-8}$ & 387& 509.2\\
    $0.002$ & $ 12.00$ & $0.90$ & $6.5431\cdot 10^{4}$ & $0.003$ & $0.007$ & $0.2909$ & $0.2906$ & $2.5136\cdot 10^{-4}$ & $3.8417\cdot 10^{-9}$ & 377& 337.2 \\
    $0.002$ & $ 12.00$ & $0.95$ & $1.1773\cdot 10^{5}$ & $0.006$ & $0.087$ & $0.3155$ & $0.3070$ & $8.4867\cdot 10^{-3}$ & $7.2089\cdot 10^{-8}$ & 410&457.9 \\
    $0.002$ & $ 15.00$ & $0.95$ & $2.8257\cdot 10^{5}$ & $0.018$ & $0.084$ & $1.3048$ & $1.3042$ & $6.2802\cdot 10^{-4}$ & $2.2225\cdot 10^{-9}$ & 424&365.2 \\
\hline
\hline
\end{tabular}
\tablefoot{Initial metallicity, mass and rotation rate (columns 1 to 3), time spent at the critical limit in Myr (column 4) and in units of the MS lifetime (column 5), central hydrogen mass fraction when the critical limit is reached for the first time (column 6), total mass lost during the whole computed evolution (column 7), mass lost through stellar winds (column 8), mass lost mechanically at the critical limit (column 9), mechanical mass-loss rate averaged over the critical-rotation period (column 10), mean equatorial velocity during the MS (column 11), and enhancement factor of the N/C mass-fraction ratio between the ZAMS and the end of the MS (column 12).}
\label{TabBeModels}
\end{table*}

\subsection{Surface velocities and abundances for very rapid rotators}

For the 9 $M_{\sun}$ models at $Z_{\sun}$, the minimum average velocity needed to reach the critical velocity is around 330 km s$^{-1}$. At $Z_\text{SMC}$ this minimum value is higher by about 20\% ($\sim 390$ km s$^{-1}$). For the 1.7 $M_{\sun}$ models, the minimum average velocity amounts to 240 and 270 km s$^{-1}$ at $Z_{\sun}$ and $Z_\text{SMC}$.

At the end of the MS phase, quite large enhancement factors of the N/C abundance ratios are obtained at the surface of the initially-rapid rotators. While at $Z_{\sun}$ it amounts to a factor of 2 for the 1.7 $M_{\sun}$ model, it amounts to a factor of 20 or more for the 9 $M_{\sun}$ model. The corresponding enhancement factors at $Z_\text{SMC}$ are between 20-40 for the 1.7 $M_{\sun}$ model and around 500 for the 9 $M_{\sun}$ model.

It is interesting to note that when these very rapid rotators evolve into the red-giant or red-supergiant stage, they keep some traces of their former very high rotation rate (see Tables 2, 3, and 4 in Paper I). This is reflected in two observable quantities: 1) their rotation rate is higher, typically the model of 2.5 $M_{\sun}$ at $Z=0.014$ with $\omega_\text{ini}=0.10$ has a rotation period $P_\text{rot}=950$ days at the end of 
the core He-burning phase, while the corresponding model with $\omega_\text{ini}=0.95$ has $P_\text{rot}<150$ days at the same evolutionary stage, {\it i.e.} it rotates at least six times faster; 2) the N/C ratio is enhanced in formerly rapid rotators, typically the 2.5 $M_{\sun}$ originating from the $\omega_\text{ini}=0.10$ model has a N/C ratio of 1.4 (in mass fraction) at the end of the core He-burning phase, while the corresponding model with $\omega_\text{ini}=0.95$ has $\text{N/C}=4.8$, {\it i.e.} more than three times higher. These differences are stronger at lower metallicities. Typically, the rotation for the 2.5 $M_{\sun}$ at $Z=0.002$ at the end of the core helium-burning phase is enhanced by a factor of 7 between the $\omega_\text{ini}=0.10$ and $0.95$ models. The N/C ratio is more than 12 times higher in the initially most rapidly rotating model than in the slower one.

From the above results, we see that it might be very interesting to study red giants in clusters that show a large number of Be stars on the MS band. From the present models, one expects that the red giants that evolved from the Be-star population present surface velocities and N/C ratios higher than normal.

\subsection{Expected number of rapidly rotating B-type stars in clusters as a function of age}

\begin{figure}
\centering{
\includegraphics[width=.45\textwidth]{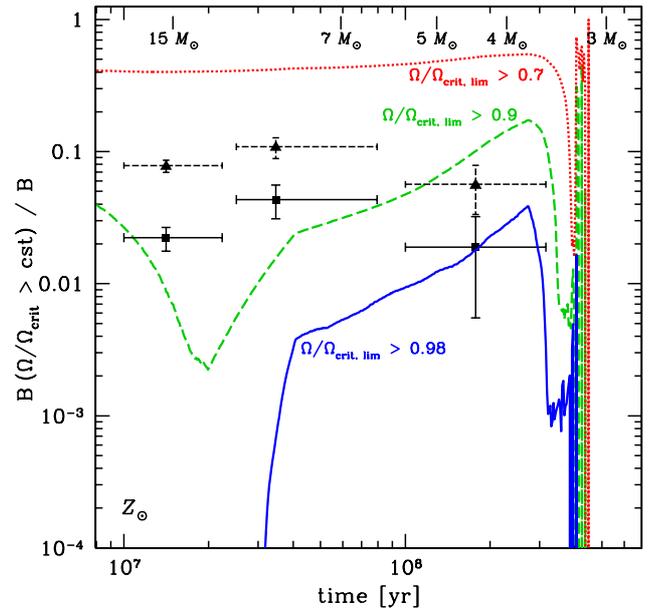}}
\caption{Evolution of the fraction of stars that rotate faster than a given limit in clusters of various ages at $Z=0.014$, taking the \citet{Huang2010a} distribution of velocities as the initial velocity distribution. The observational points represent the observed Be/(B+Be) ratio \citep{McSwain2005b}. The squared symbols (solid error bars) are for confirmed Be stars, and the triangle symbols (dashed error bars) are for all Be-type stars. The ticks at the top of the plot show the typical lifetime of stars with various initial masses.}
\label{z014H2010}
\end{figure}

\begin{figure}
\centering{
\includegraphics[width=.45\textwidth]{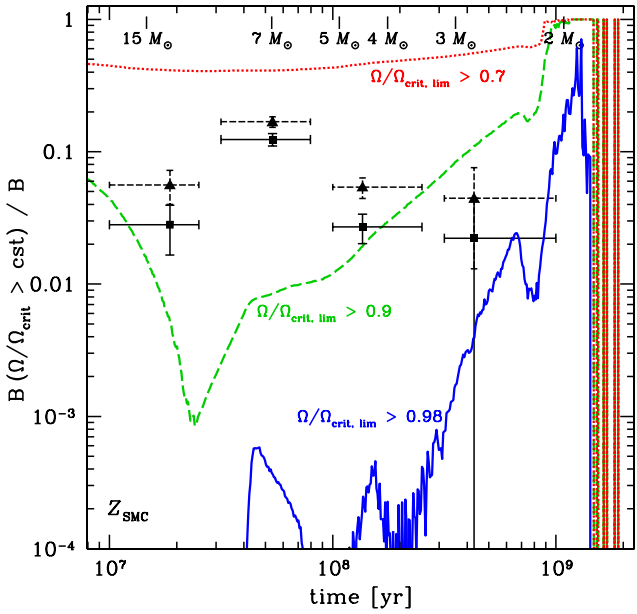}}
\caption{Same as Fig.~\ref{z014H2010} for $Z=0.002$. Observed Be/(B+Be) ratio extracted from \citet{Martayan2010a}, following the same procedure as in \citet{McSwain2005b}. We consider the following age beams: $7 \le \log(t) < 7.5$, $7.5 \le \log(t) < 8.0$, $8.0 \le \log(t) < 8.5$, and $\log(t) �\ge 8.5$. The error bars for the fraction of Be-type stars are Poisson statistic error bars.}
\label{z002H2010}
\end{figure}

The fraction of stars with surface velocities above a given limit varies in clusters of various ages, and it can be estimated from the computed grids. This is shown in Figs.~\ref{z014H2010} and \ref{z002H2010} for solar and SMC metallicity. The populations were computed as follows. First, the mass range ($1.7\, M_{\sun}$--$15\, M_{\sun}$) and initial rotation-rate range ($0 \le \omega_\text{ini} \le 1$) were discretised in several intervals (1000 mass intervals and 100 velocity intervals), dividing the plane $M_\text{ini}$--$\omega_\text{ini}$ into 100 000 cells. For each cell, we extracted the expected number of stars (normalised to an initial population of 1), according to an initial mass function \citep{Salpeter1955}, and to an initial rotation-rate distribution \citep{Huang2010a}. Each cell thus defines initial conditions in $Z$, $M$, and $\omega$. We built an evolutionary track for these initial conditions interpolating between our computed tracks. At each time step, we checked the spectral type of the star in each of the cells, as well as its current rotation rate. If this rotation rate was higher than a given constant ($0.70$, $0.90$, or $0.98$), we added the number of stars in the cell to the total number of stars of a given spectral type that rotates more rapidly than the given constant. The population was normalised with respect to the total number of B stars (including Be stars) at the same time\footnote{We consider a star to be a B-type star if $10,000\,[\text{K}] < T_\text{eff} < 30,000\,[\text{K}]$.}. We assumed that the distribution of initial velocities given by \citet{Huang2010a} corresponds to the distribution of  velocities on the ZAMS, {\it i.e.} before the slowing down due to the activity of the meridional currents described above in Sect.~\ref{secprop}. 
We emphasise here once more that an appropriate account of the way the mass and angular-moment content of the star builds up during the accretion phase may change this picture.
This will be studied in detail in a forthcomming paper (Haemmerl\'e et al. in preparation).

Under the previous hypotheses we find that at $Z_{\sun}$, the fraction of stars with $\omega > 0.98$ 
(very rapid rotators hereafter, blue solid line in Fig.~\ref{z014H2010}) becomes greater than 1\% after 100 Myr, 
when the initially rapidly rotating stars with masses between 4 and 5 $M_{\sun}$ approach the critical limit. 
The fraction of very rapid rotators then increases with the age of the cluster until a maximum at around 4\% is reached at an age of about 300 Myr.

The curve for $\omega > 0.90$ (green dashed) presents a local minimum around 20 Myr. 
This reflects the fact that in the first part of the evolution of our lower initial mass model, the $\omega$ ratio decreases.
 Therefore this first behaviour is quite dependant on the initial conditions chosen. 
Then the curve follows a similar trend as the one for $\omega > 0.98$, but shifted at higher values. 
In contrast, the line corresponding to $\omega > 0.70$ (red dotted) is almost constant up to about 400 Myr.

Comparing the values obtained at $Z=0.014$ (Fig.~\ref{z014H2010}) and $0.002$ (Fig.~\ref{z002H2010}), a few interesting remarks can be made:
\begin{itemize}
\item for most of the age interval between 30 and 400 Myr, the fraction of stars with $\omega > 0.98$ at a given age is greater at $Z_{\sun}$ than at $Z_\text{SMC}$.
This occurs because 4 and 5 M$\odot$ models at Z$_{SMC}$ do not attain values of $\omega$ close to 1 (See also Table 1):
even though the rotational velocities do not differ significantly between SMC and solar metallicities for the mentioned masses, these models do 
not approach the critical limit because the stars are more compact and their critical velocities are higher at SMC metallicities.

\item Stars with $\omega > 0.98$ are found in older clusters at $Z_\text{SMC}$ than at $Z_{\sun}$. 
This is because at SMC metallicity, at ages older than 1 Gyr only the lower mass rapidly rotating B-type models are still in their main sequence. 
At this age and metallicity, all other models with masses higher than 1.7 $M_{\sun}$ and those of the same mass but with medium or low rotation, 
will have ended their core-hydrogen burning. 
The enhancement in the main-sequence lifetime for 1.7 M$\odot$ rapidly rotating models occurs because their central temperature attains values for which the
 pp-chain is almost entirely responsible for the energy production during a long part of the MS phase \citepalias[see discussion and Figure 5 in][]{Georgy2012c}.
 
\item For stars with $\omega > 0.90$, the differences between $Z_{\sun}$ and $Z_\text{SMC}$ are less marked, except that at the $Z_\text{SMC}$ some older clusters show a significant fraction of $\omega > 0.90$ stars and that at very high ages (above the billion years), a great fraction of the B-type stars may be rapid rotators. This is expected because only the most rapidly rotating models are still B-type stars on the MS band at these ages. They are some kind of blue stragglers\footnote{At $Z=0.002$ we slightly underestimate the number of B-type stars since some of them would come from stars with a mass lower than 1.7 $M_{\sun}$, and thus we overestimate the Be/(B+Be) ratio.}.
\end{itemize}

Implications of these results concerning the question of the Be phenomenon are discussed in Sect.~\ref{secbe}.

\section{Mechanical mass loss at the critical limit \label{secmec}}

In this section, we explain how the mechanical mass losses (see columns 9 and 10 in Table~\ref{TabBeModels}) were computed.
We want to stress that in the frame of the hypotheses  described below, we obtain an outbursting behaviour for the mass loss.
However, this outbursting behaviour results from our numerical procedure and does not come from ab initio physical principles. In this section we
 explain in detail how the mechanical mass losses were computed and study how the amount of mass lost depends on the time step and on $\omega_{\rm max}$.

\begin{figure}
\centering{
    \includegraphics[width=9.0cm]{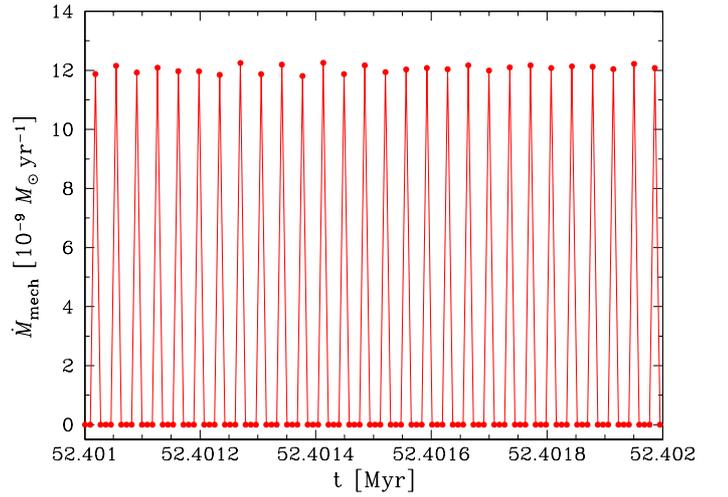}}
    \caption{Evolution of the mechanical mass-loss rate as a function of time for the  7 $M_{\sun}$ stellar model at solar metallicity with a time step of about nine years (case $n=105$, see text). Each point represents the mass loss rate in one time step. The quiescent phase between two mass-ejection episodes lasts for about 40 years.}
    \label{afer}
\end{figure}

The general principle is the following: 
owing to internal transport of angular momentum and to the stellar evolution itself, it may occur, depending on the initial rotation, that the stellar surface reaches (and even surpasses) the critical velocity (or rather $\omega_\text{max}$ in our computation, see Sect.~\ref{secmod}) during the evolution. In this case, the excess angular momentum is removed from the outer layers in the form of an additional equatorial mass loss on a dynamical timescale, {\it i.e.} within one time step. This brings the surface velocity of the star to below 0.99 $\Omega_\text{crit}$. Ideally, the stellar surface should rotate at 0.99 $\Omega_\text{crit}$, but it does 
not, because of discretisation. 
The transport of angular momentum by meridional currents and shears will then accelerate the surface until the limit is crossed again and some more mass is shed. For illustration purposes, the evolution of the mass loss obtained this way is shown in Fig.~\ref{afer} during a period spanning about 1000 years for our 7 $M_{\sun}$ stellar model at solar metallicity. We see a very regular pattern where the quiescent phase between two mass-ejection episodes lasts for about 40 years, and the mass ejections are of the order of 10$^{-8}$ $M_{\sun}$ yr$^{-1}$. For this model, the radiative winds are negligible and thus the only mass-loss mechanism taking place is mechanical mass loss, as described above.

In the following, we explore the dependency on the results on the time step used for the calculations and on the parameter $\omega_\text{max}$.

\subsection{Discussion of the choice of the time step}

First, we briefly examine how the choice of the time steps modifies the amplitude of the mechanical mass-loss rates. Depending on 1) the duration of the time steps, 2) the equatorial velocity of the star just before becoming overcritical, and 3) the efficiency of the angular-momentum transport, the surface velocity during the next time step may surpass the critical limit to a lesser or greater extent. In some cases, very weak mechanical mass-loss episodes will occur (typically when the amplitude of the evolution of  $V_\text{eq}$ beyond the critical limit will just graze the limit), while in some other cases, very strong mass-loss episodes will occur, when $V_\text{eq}$ strongly surpasses the limit. Therefore our numerical method leads to some scatter in the mass-loss rates. In Fig.~\ref{afer}, the interval shown is too short for this scatter to appear.

To study the impact of the choice of the time step on the outputs of our models, we computed different $7\, M_{\sun}$ models at $Z=0.014$ with $\omega_\text{ini}=0.95$. The first column of Table~\ref{TabTimestep1} indicates the value of $n$, characterising the different time steps $\Delta t = \Delta t_\text{standard}/n$, where $\Delta t_\text{standard}=942\,\text{yr}$. The last two columns show the total mass lost and the mean mass-loss rate during the considered period, which in the present case corresponds to a period of 250 000 yr beginning when the age of the star is $5.21\cdot10^7$ yr. When the time step becomes lower than about 100 years, these two quantities stabilise around $7.3\cdot10^{-4}\ M_{\sun}$ for the total mass lost mechanically during that period and towards $2.9-3.0\cdot10^{-9}\ M_{\sun}$ per year for the time-averaged mass-loss rates. This shows that below a given value for the time step, the results are no longer varying. We also see that taking a time step greater by an order of magnitude does not change the results by an order of magnitude: the difference amounts at most to a factor of two in the range of the time steps studied here.

\begin{figure}[h]
\centering{
    \includegraphics[width=9.0cm]{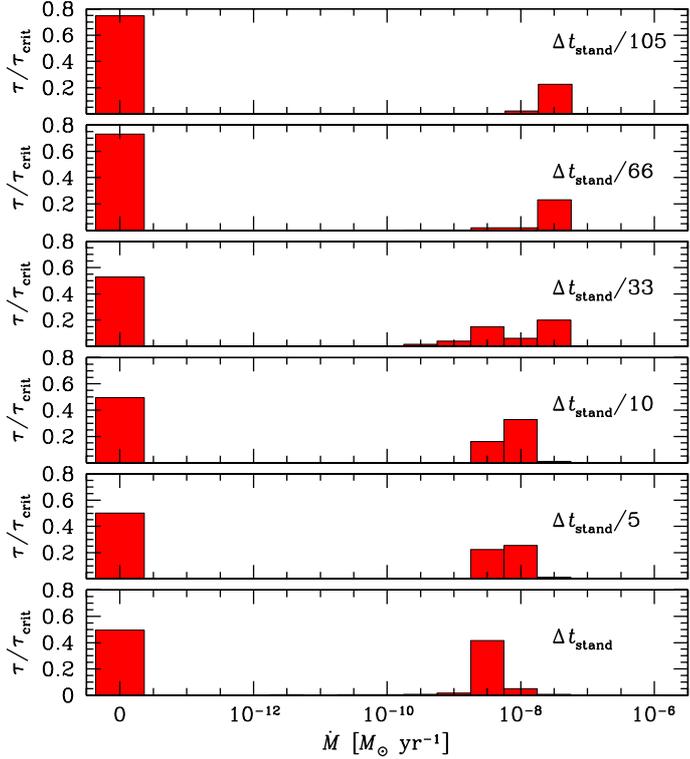}}
    \caption{Fraction of time spent with a given mass-loss rate for the 7 $M_{\sun}$ models at $Z=0.014$ with various time steps.}
    \label{fraction}
\end{figure}

Figure ~\ref{fraction} shows the fraction of time spent near the critical limit with zero and non-zero mass-loss rates for our 7 $M_{\sun}$ model at $Z=0.014$. Depending on the time step used, the fraction of the time spent away from the critical limit, {\it i.e.} when the star is rotating fast but does not lose mass through mechanical mass loss, amounts between 50 and 80\% of the critical period. Shorter time steps tend to increase the duration of the null mechanical mass-loss period and to increase the mass-loss rates in the period when mass ejections occur. The differences become small when time steps shorter than 100 yr are used.

As a conclusion, we see that the scatter brought to the results by a change in the time step is not dramatic. The adoption of the longest time steps does not change the results by more than a factor of two with respect those obtained with the shorter time steps, and also 
allows for a reasonable computational time for these extended grids.

To support these numerical results, we compared them with a simple analytical estimation of the mechanical mass loss rate.
Asuming that the star rotates as a solid body, that the rotational velocity change is very small when the star rotates close to the critical limit,
and also that the time variation of the stellar angular momentum of the star is due to the loss of matter through the equator, 
  we obtain simply that $\dot{\text M} = \dot{\text I}/R_{eq}^2$. This shows that the mechanical mass loss rate when the star rotates at the critical 
limit depends basically on the evolution of the momentum of inertia mass and the equatorial radius, which both depend on the stellar mass. 
With this approximation, within the same interval of time as described above, we obtain a
mass loss rate of $1.5\times10^{-9}$ for a 7 $M_{\sun}$ model at Z=0.014, which nicely agrees with the numerical results.

\begin{table}
\begin{center}
\caption{Time-step dependence of the total mass lost mechanically and of the mean mechanical mass-loss rates (in the 250'000 yr interval).}
\label{TabTimestep1}
\begin{tabular}{lccccc}
\hline\hline
\rule[0mm]{0mm}{3.5mm}$n$ & $\Delta t$ & $\Delta M$ & $<\dot M>$                       \\
       &\rule[-1.5mm]{0mm}{2mm} yr          & $10^{-4}\ M_{\sun}$ &$10^{-9}\ M_{\sun}\,\text{yr}^{-1}$ \\ 
       \hline
     \rule[0mm]{0mm}{3mm}1 & 942         & 4.087     & 1.635                         \\
     5 & 188         & 6.684     & 2.674                         \\ 
   10 &   94         & 7.289     & 2.916                         \\
   33 &   28         & 7.533     & 3.013                         \\
   66 &   14         & 7.258     & 2.903                         \\
  105 &    9         & 7.290    & 2.916                         \\
\hline\hline
\end{tabular}
\end{center}
\end{table}

\subsection{Choice of the value of $\omega_\text{max}$ at which mechanical mass loss occurs\label{SSSOmegaMax}}

As emphasised above, because of the numerical difficulties of computing the stellar evolution when the star is at the critical limit, we need to define a value for the highest fraction of the critical velocity allowed, which we call $\omega_\text{max}$ (see Sect.~\ref{secmod}) above which the mechanical mass loss sets in.

To check to what extent the choice of $\omega_\text{max}$ affects the evolution of the star, we computed $7\, M_{\sun}$ models at solar metallicity with $\omega_\text{ini}=0.95$, setting  $\omega_\text{max}$ equal to 0.99, 0.95, and 0.90 just after the readjustment phase at the very beginning of the MS. The results are presented in Table \ref{tabla7}. The first column gives $\omega_\text{max}$, the second and third one the MS lifetime and the fraction of the MS lifetime spent near the critical limit; the total mass lost and the time-averaged mass-loss rate over the critical period are given in columns 4 and 5.

We see that the lower $\omega_\text{max}$, the longer are the MS lifetimes, and the higher it is the ratio $\tau_\text{crit}/\tau_\text{MS}$. This is a consequence of the 
surface braking by the mechanical mass loss, which occurs earlier for the models with the lower values of $\omega_\text{max}$, producing a stronger internal mixing in the star. This mixing brings fresh hydrogen to the core, increasing its size, slightly increasing its luminosity, and also the MS lifetime. The total mass lost and the mechanical mass-loss rates are also higher for lower $\omega_\text{max}$. This is expected, since a lower value of $\omega_\text{max}$ implies an earlier entrance into the critical regime, more mass removed when the limit is crossed, and a higher amount of total mass lost. Typically, passing from $\omega_\text{max}=0.99$ to $0.95$ already increases the total mass lost by more than a factor of two. 
Because both the total mass lost and the duration of the critical period increase in a similar way when $\omega_\text{ini}$ decreases, 
the averaged mass loss rate remains nearly constant. 

\begin{figure}
\includegraphics[width=.45\textwidth]{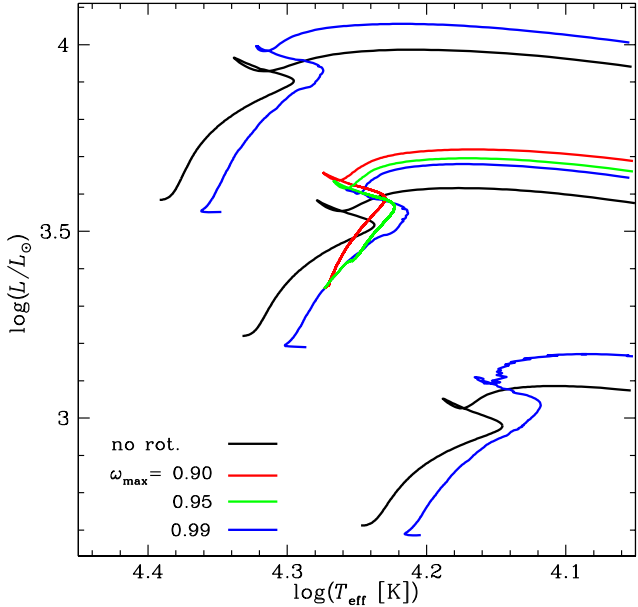}
\caption{HR diagram for $7\, M_{\sun}$ models at solar metallicity computed with different values of $\omega_\text{max}$ (blue: $0.99$, green: $0.95$, and red: $0.90$). As a comparison $5$ and $9\, M_{\sun}$ evolutionary tracks are also plotted.}
\label{omecrit}
\end{figure}

Figure \ref{omecrit} shows the HR diagram for the $7\, M_{\sun}$ models computed with the different values of $\omega_\text{max}$, and as a comparison the evolutionary tracks corresponding to the non-rotating case and the evolutionary tracks corresponding to $\omega_\text{max} = 0.99$, for $5$, $7$ and $9\, M_{\sun}$. The differences in $\log(L)$ and $\log(T_\text{eff}$) have different causes. As mentioned above, the luminosity shift is due to the larger core of the model with the lower $\omega_\text{max}$. The $T_\text{eff}$ shift has a geometrical origin. For a given luminosity, the surface of a star rotating with $\omega=0.99$ is $1.16$ times larger than the surface of a star with $\omega=0.90$, because of the deformation of the shape of the surface \citep[see][]{Georgy2011a}. Because the mean $T_\text{eff}$ is deduced according to the Stefan-Boltzmann law $L\sim \Sigma T_\text{eff}^4$ (with $\Sigma$ the stellar surface), this produces a shift of $0.016$ dex towards the red in the HR diagram.

In the rest of the paper, we discuss the results obtained with  $\omega_\text{crit}=0.99$.
To conclude this section we stress two points:
\begin{enumerate}
\item Reality may be quite different from the situation described above. It might occur, for instance, that mechanisms other than critical rotation are involved in
producing mass ejection. These additional mechanisms may allow the loss of matter well before the surface rotational velocity becomes critical.
\item While the present approach is of course probably too schematic, it can at least provide a reasonable estimate of the mass lost mechanically when the following hypotheses are made: mechanical mass loss only occurs when the velocity is higher than a fixed limit and the quantity of mass lost is entirely determined by the necessity of the star to shed overcritically rotating material. 
\end{enumerate}

\begin{table}[t]
\caption{Properties of the $7\, M_{\sun}$ stellar models at $Z=0.014$ computed with various values of $\omega_\text{max}$.}
\begin{center}
\begin{tabular}{ccccc}
\hline\hline
\rule[0mm]{0mm}{3mm}$\omega_\text{max}$ & $\tau_\text{MS}$ & $\tau_\text{crit}/\tau_\text{MS}$ & $\Delta M_\text{mech}$ & $\dot{M}$ \\
 \rule[-1.5mm]{0mm}{5mm} & $10^7\,\text{yr}$ & & $M_{\sun}$ & $M_{\sun}\,\text{yr}^{-1}$ \\
\hline
\rule[0mm]{0mm}{3mm}0.90 & 5.573 & 0.476 & 0.043 & $1.62\cdot 10^{-9}$\\%
0.95 & 5.398 & 0.269 & 0.021 & $1.42\cdot 10^{-9}$\\%
0.99 & 5.336 & 0.115 & 0.009 & $1.35\cdot 10^{-9}$\\%
\hline\hline
\end{tabular}
\end{center}
\label{tabla7}
\end{table}

\subsection{Radiative and mechanical mass loss}

In Table \ref{TabBeModels}, column 10,  we show the mass-loss rates averaged over the period during which the star evolves near the critical limit for stars of different stellar masses and metallicities. 

The total mass and angular momentum lost mechanically are quite small with respect to the total mass and angular-momentum content. Typically, for the 15 $M_{\sun}$ model at $Z=0.002$, the total mass lost mechanically corresponds to 0.005\% of the initial mass. This means that this mass-loss mechanism has only very little influence on the evolution of the star.

For a given metallicity, the mean mechanical mass-loss rates are higher for higher stellar masses, 
As mentioned in subsection 4.1, this dependence can also be deduced from simple analytical calculations:
the mass-loss rates depend on the evolution of the stellar momentum of
inertia and on the equatorial radius of the stars, and both quantities depend on the initial stellar mass.
For a given mass, we find that the mechanical mass-loss rates are in general higher at lower $Z$. In the high mass star range, this can be explained easily: 
indeed, in the high-mass range, the stronger radiative winds of higher $Z$ remove mass and angular momentum at the surface, keeping the star more easily away from the critical limit and thus reducing the need for a mechanical mass loss. In the low-mass range, the behaviour is more complex, which is the result of counteracting effects. First, stars are more compact at low $Z$, so the extent of the region over which angular momentum must be transported is smaller; second, this compactness leads to weaker meridional currents \citep{Maeder2001a}, which reduces the transport efficiency. The present results seem to imply that in the metallicity range presented here, the first effect overcomes the second. From the above, we conclude that 
our mechanical mass loss rates increase with mass, as is the case of mass loss rates via radiatively line-driven winds, 
but  mechanical mass loss is higher at lower Z, in contrast to what happens for stellar winds.

An interesting point to be highlighted here is the impact of the mass loss on the angular-momentum content of the stars. The total initial angular momentum of models that rotate with different $\omega_\text{ini}$ remains within an order of magnitude for a given stellar mass. Along the main-sequence evolution, the angular momentum of the star will be modified by the effect of stellar winds (for stars with masses larger than $7\, M_{\sun}$) and by the mechanical angular-momentum loss, for stars that reach the critical limit.  The stellar winds dominate the high-mass range, driving the loss of about 30\% of the initial angular momentum at $Z=0.002$. The mechanical mass loss reduces the angular momentum at most up to 10\%, but is very effective in injecting angular momentum into the circumstellar environment, probably forming a circumstellar disk.

\subsection{Polar and equatorial mass fluxes\label{secpoleq}}

If mass loss occurs only via radiatively driven winds, the equatorial to polar mass-flux ratio depends only on the value of $\omega$ \citep{Maeder2002a,Georgy2011a} and basically reflects the temperature contrast between the pole and equator described by the von Zeipel's theorem. For a star rotating at  $\omega=0.99$ the equatorial to polar mass-flux ratio is therefore predicted to be at most $0.15$, if there is no mechanical mass loss \citep{Georgy2011a}. We see in Table~\ref{fluxes} that for the models with 7 and 9 $M_{\sun}$, the mean mass flux is higher at lower metallicities. This might be surprising at first sight since a stronger mass loss by radiative winds is expected at high metallicity because of the metallicity dependence of the line-driven winds (which scale as $\sim Z^{-0.5}$). However, polar winds occur only when the surface velocity is close to the critical limit and the evolutionary stage when this occurs depends on the metallicity, and at low metallicity the critical limit is reached closer to the end of the MS, where the radiative winds are stronger due to the higher luminosity of the star. 

When mechanical mass loss occurs, the mass flux at the equator consequently becomes much greater than the mass flux at the pole (enhancement factors of up to 60) in most cases. The only exceptions occur for the highest initial masses in each metallicity. In that case, the radiative winds are stronger and thus remain of the same order of magnitude (or even stronger in one case) than the equatorial mechanical mass flux.

\begin{table*}[t]
\caption{Mean mass fluxes during the critical-rotation phase for stars with $\Omega_\text{ini}/\Omega_\text{crit}=0.95$.}
\label{fluxes}
  \begin{center}
\begin{tabular}{ccccccc}
\hline\hline
\rule[0mm]{0mm}{3mm}$Z$&$M$&Flux$_\text{pol}$& Flux$_\text{eq}$&Flux$_\text{mech}$&Flux$_\text{eq}$/Flux$_\text{pol}$&Flux$_\text{mech}$/Flux$_\text{pol}$\\
\rule[-1.5mm]{0mm}{3mm}&$M_{\sun}$&$M_{\sun}\,\text{yr}^{-1}\,\text{sr}^{-1}$ & $M_{\sun}\,\text{yr}^{-1}\,\text{sr}^{-1}$& $M_{\sun}\,\text{yr}^{-1}\,\text{sr}^{-1}$& \\
\hline
\rule[0mm]{0mm}{3mm}$0.002$ & $7$ & $1.854\cdot10^{-11}$ & $2.790\cdot10^{-12}$ & $4.711\cdot10^{-10}$ & 0.15 & $25.4$\\
$0.002$ & $9$ & $2.242\cdot10^{-10}$ & $3.372\cdot10^{-11}$ & $2.658\cdot10^{-9}$  & 0.15  &$11.9$\\
$0.002$ &$12$ & $1.609\cdot10^{-9}$  & $2.420\cdot10^{-10}$ & $1.148\cdot10^{-8}$  & 0.15  &$7.13$\\
$0.002$ &$15$ & $4.657\cdot10^{-9}$  & $7.005\cdot10^{-10}$ & $3.533\cdot10^{-10}$ & 0.15 &$0.23$\\
$0.006$ & $7$ & $1.296\cdot10^{-11}$ & $1.950\cdot10^{-12}$ & $7.878\cdot10^{-10}$ & 0.15 &$60.8$\\
$0.006$ & $9$ & $2.030\cdot10^{-10}$ & $3.053\cdot10^{-11}$ & $9.247\cdot10^{-10}$ & 0.15 &$4.02$\\
$0.006$ &$12$ & $2.323\cdot10^{-9}$  & $3.496\cdot10^{-10}$ & $3.899\cdot10^{-9}$  & 0.15  &$1.68$\\
$0.014$ & $7$ & $1.003\cdot10^{-11}$ & $1.509\cdot10^{-12}$ & $2.148\cdot10^{-10}$ & 0.15 &$21.4$\\
$0.014$ & $9$ & $1.680\cdot10^{-10}$ & $2.528\cdot10^{-11}$ & $6.685\cdot10^{-10}$ & 0.15 &$3.98$\\
\hline\hline
\end{tabular}
\tablefoot{Mean polar and equatorial mass fluxes due to radiative winds (columns 3 and 4 respectively), mechanical mass flux (column 5), polar to equatorial mass-flux ratio due to anisotropic winds (column 6), and polar to mechanical mass-flux ratio (column 7). We only consider here models for which we included a radiative mass-loss prescription during the MS ($M_\text{ini}\ge7\,M_{\sun}$).}
\end{center}
\end{table*}

\section{Decretion disks \label{secdisks}}

\subsection{Estimation of the physical properties}

\citet{Struve1931a} first proposed the existence of a decretion disk formed from an equatorial-mass ejection episode by a rapidly rotating star. Since then, the existence of Keplerian rotating disks has been increasingly supported \citep[e.g.][]{Porter2003a,Carciofi2011a}.

In the previous section, we showed that the surface velocity remains at around the critical limit from the first moment it reaches it until the end of the MS phase, and that mass ejections in the frame of the hypotheses made here occur as intermittent outbursts. The ejected mass launched into an equatorial disk around the star explains the observed Balmer emission lines of Be stars. 

We study here the physical characteristics of a stationary disk that can be deduced from the developments made by \citet{Krticka2011a}, who solved the hydrodynamic equations in cylindrical coordinates to obtain the disk structure. 
\citet{Krticka2011a} provided parametrised expressions for the mass-loss rate as a function of the angular-momentum-loss rate and for the outer radius of the disk.  We used these simple parametrised expressions to study how the disk properties change along the critical-rotation phase.

There are three free parameters in this disk model: two of them, $p$  and $T_0$, arise from the assumption that the temperature in the disk varyies as a power-law in radius, $T=T_0(R_\text{eq}/r)^p$, where $R_\text{eq}$ is the equatorial radius of the star. The third free parameter is the viscosity term $\alpha$ of \citet{Shakura1973a}, which intervenes in the momentum equation (tangential component). This viscosity couples the inner parts of the disk  to the central star and may be generated by the presence of small magnetic fields \citep{Balbus1991a}.

The structure of the disk and our stellar model are connected through three variables: 1) $\dot{\mathcal{L}}$, which is the quantity of angular momentum that the star must lose per unit of time to maintain the surface velocity at the critical limit (at $\omega_\text{max} = 0.99$); 2) $R_\text{eq}$, the equatorial radius of the star; and 3) $V_\text{K}(R_\text{eq})$,  the Keplerian velocity at the equator of the star (see more details below).

Part of the excess angular momentum that the star has to lose is lost through radiative stellar winds. To calculate the angular-momentum-loss rate by this process, $\dot{\mathcal{L}}_\text{wind}$, we inserted the radiative mass-loss rates $\dot M$ given by our evolutionary models and assumed that this mass is lost isotropically at the mean stellar radius given by $R_\text{mean}=\sqrt{L/4\pi \sigma T_\text{eff}^4}$, which gives $\dot{\mathcal{L}}_\text{wind}=2/3 \dot M \Omega R_\text{mean}^2$. By doing so, we  neglected the wind anisotropies and thus  slightly overestimate the loss of angular momentum due to this process. However, except for the 15 $M_{\sun}$ stellar model at $Z=0.002$, the stellar winds remove a negligible fraction of the mass and of the angular momentum even when the effects of the stellar winds are overestimated. The remaining angular-momentum-loss rate, $\dot{\mathcal{L}}-\dot{\mathcal{L}}_\text{wind} $, is transferred to the disk ($\dot{\mathcal{L}}$ hereafter for clarity).

In a stationary situation, the disk will lose angular momentum at the same rate. Therefore, knowing how and where the disk loses this angular momentum will set up the disk structure corresponding to the considered physical situation. The disk can lose angular momentum through two different channels:
\begin{enumerate}
\item through disk winds ($\dot{\mathcal{L}}_\text{dw}$). The value for the maximum $\dot{\mathcal{L}}_\text{dw}$ is obtained from expression (26) in \citet{Krticka2011a}, which corresponds to the highest disk-wind angular-momentum-loss rate.\footnote{If this highest $\dot{\mathcal{L}}_\text{dw}$ were be greater than the angular momentum carried by the disk outflow, $\dot{\mathcal{L}}-\dot{\mathcal{L}}_\text{wind}$, the extension of the disk would be set by the radius $R_\text{out}$, given by \citet{Krticka2011a} such that $\dot{\mathcal{L}}_\text{dw}(R_\text{out})=\dot{\mathcal{L}}-\dot{\mathcal{L}}_\text{wind} $. This never happens in our models, see text.} In all our models the highest $\dot{\mathcal{L}}_\text{dw}$ is negligible with respect to $\dot{\mathcal{L}}$ (compare column 5 with column 3 in Table~\ref{tabla0}), therefore we neglect the effect of a disk wind hereafter;
\item through disk mass loss ($\dot{\mathcal{L}}_\text{disk}$). Since this is the only active process in most cases, one has $\dot{\mathcal{L}}=\dot{\mathcal{L}}_\text{disk}$.
\end{enumerate}

The quantity $\dot{\mathcal{L}}$ is given by our evolutionary calculations, while the value of $\dot{\mathcal{L}}_\text{disk}$ is obtained by \citet{Krticka2011a}. These authors showed that at a radius $R_\text{crit}$ the radial velocity of the matter in the disk becomes equal to the local sound velocity, $a=\sqrt{kT/\mu m_\text{H}}$, where $\mu$ is the mean molecular weight and $m_\text{H}$ the mass of a hydrogen atom. Matter flowing beyond $R_\text{crit}$ removes angular momentum from the star-disk system. 

Accounting for the fact that at large distances the disk is not rotating at the Keplerian velocity,
$$\dot{\mathcal{L}}_\text{disk}= \frac{1}{2} \dot M_\text{disk} V_\text{K}(R_\text{crit}) R_\text{crit},$$
where $\dot M_\text{disk}$ is the mass lost by the disk, $V_\text{K}(R_\text{crit})$, the Keplerian velocity at $R_\text{crit}$, {\it i.e.} $V_\text{K}(R_\text{crit})= \sqrt{GM/R_\text{crit}}$. The equality between $\dot{\mathcal{L}}$ and $\dot{\mathcal{L}}_\text{disk}$ allows us to determine  $\dot M_\text{disk}$, provided $R_\text{crit}$ is known. 

\citet{Krticka2011a} proposed an approximation of the critical radius, which can be considered as the outer radius of the disk:
\begin{equation}
\frac{R_\text{out}}{R_\text{eq}}\sim\frac{R_\text{crit}}{R_\text{eq}}=\left[ \frac{3}{10+4p} \left( \frac{{\rm \upsilon_{K}}(R_\text{eq})}{a(R_\text{eq})} \right) ^{2} \right]^{\frac{1}{1-p}},
\end{equation}
where $a(R_\text{eq})$ is the sound speed evaluated at $R_\text{eq}$.

At this point, we have three interesting quantities that characterise the disk: its radial extension, $R_\text{crit}$ (defined here as its extension up to the point when angular momentum is no longer conserved), the rate of angular-momentum loss, $\dot{\mathcal{L}}_\text{disk}$, and of mass loss, $\dot M_\text{disk}$.

We easily obtain two more interesting quantities, the diffusion time and the mass of the disk. The characteristic timescale for the evolution of the disk is given by the viscous diffusion time,
\begin{equation}   
t_\text{diff}= \frac{R_\text{crit}^2}{\nu}, 
\end{equation}
where $\nu\,=\,\tilde{\alpha}\,a\,H\,$ is the diffusivity of the disk, with $\tilde{\alpha}$ the viscosity parameter, and $H$ the disk scale height, $H=a(R)\,R^{3/2}/(G\,M)^{1/2}$, where $G$ is the gravitational constant and $M$ the mass of the star. Evaluating $H$ at $R_\text{crit}$ and assuming an isothermal disk leads to 
\begin{equation}   
t_\text{diff}= \frac{\sqrt{G\,M\,R_\text{crit}}}{\tilde{\alpha} a^2}.
\end{equation}

From the diffusion timescale and mass-loss rate of the disk, we can estimate the mass of the decretion disk. It is obtained by multiplying the mass-loss rate of the disk by the diffusion timescale:
\begin{equation} 
M_\text{disk}= \frac{2\,\dot{\mathcal{L}}}{\sqrt{{\rm G}\,M\,R_\text{crit}}}\cdot\frac{\sqrt{G\,M\,R_\text{crit}}}{\tilde{\alpha} a^2}=\frac{2\,\dot{\mathcal{L}}}{\tilde{\alpha} a^2}.
\end{equation}
The mass of the viscous decretion disk does not depend on the outer radius $R_\text{crit}$ but is set by the angular-momentum-input rate in the disk when the stellar surface reaches the critical limit.

The disk properties are computed assuming an isothermal disk ($p=0$) of temperature $T_ 0=2/3\ T_\text{eff}$ (which are the disk parameters frequently used in the literature to represent Be-star disks),  and $\alpha$=1 as proposed by \citet{Carciofi2012a}.

\subsection{Results from the stellar models}

 \begin{figure}
\centering{
    \includegraphics[width=.5\textwidth]{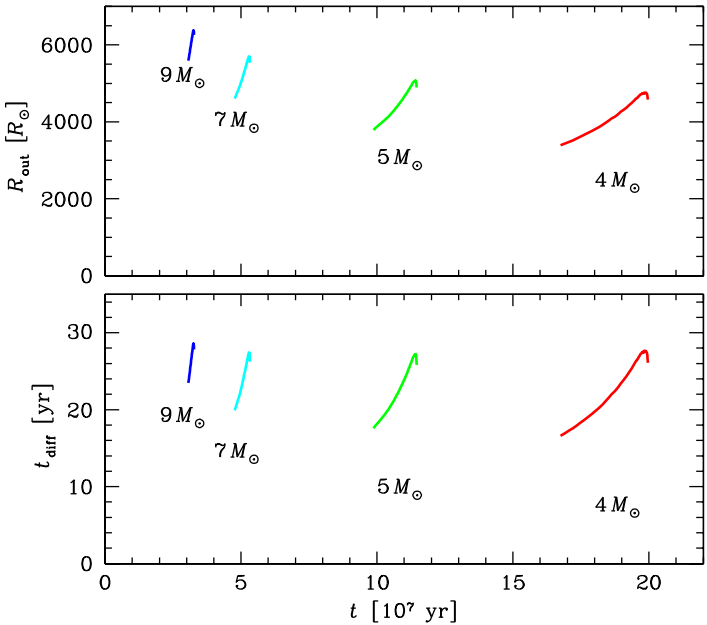}}
    \caption{Evolution of the radius and diffusion time of the disk as a function of time for critically rotating models with an initial mass between 4 and 9 $M_{\sun}$ at solar metallicity.}
    \label{evdisk1}
\end{figure}

In Table~\ref{tabla0}, we give the properties of the disks that are obtained by applying the equations of the previous section to our stellar models with $\omega_\text{ini}=0.95$. The indicated quantities are obtained by computing their time-averaged value throughout the critical period to remove the scatter due to the outbursting nature of the process.

\begin{figure}
\centering{
    \includegraphics[width=.5\textwidth]{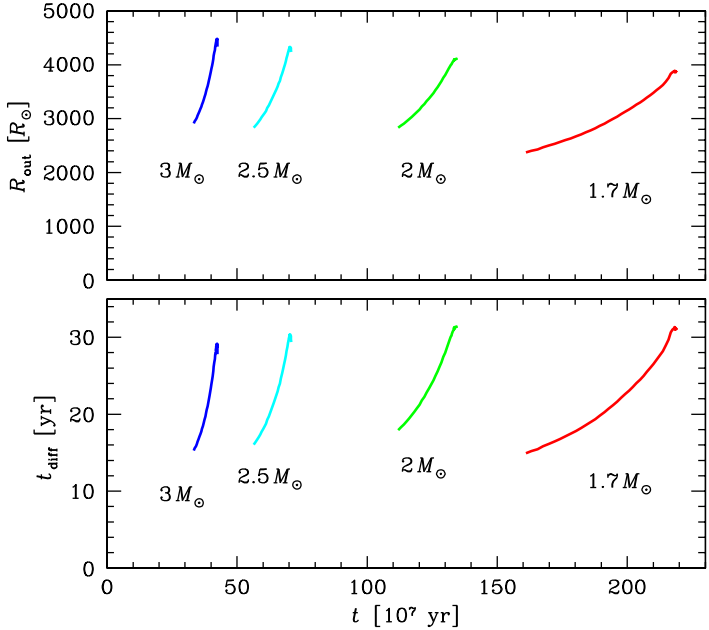}}
    \caption{Same is in Fig.~\ref{evdisk1} for an initial mass between 1.7 and 3 $M_{\sun}$.}
    \label{evdisk2}
\end{figure}

We note the following points:
\begin{itemize}
\item The mass-loss rates via the viscous decretion disk, $\langle\dot{M}_\text{disk}\rangle$, are much lower than the mass-loss rates given in Table~\ref{TabBeModels}, $\dot{M}_\text{mech, mean}$, corresponding to the mass lost at the equator of the star: in the disk, the matter is released at a farther distance, and thus a lower mass loss is needed to remove a given amount of angular momentum. The mass-loss rates from the disk are about a factor of 10 below the mass-loss rates at the surface of the star \citep{Krticka2011a}. 
\item The masses of the disks are quite small. Values between $9.4\cdot10^{-12}$ and $1.4\cdot10^{-7}\ M_{\sun}$ (3$\cdot10^{-6}$ to $0.047$ times the mass of the Earth) are obtained. These quantities are strongly dependent on the disk parameters $p$, $T_0$, and $\alpha$.
\item Under the hypotheses considered, the extension of the disk amounts to a few thousands of solar radii (between about 10 and 30 Astronomical Units). Expressed in terms of stellar radii, the extension amounts to some hundreds of stellar radii. 
\item The diffusion time is very short, only a few decades. This is much too short for any instability in the disk to set up and to form something like an asteroid belt!
\item At a given metallicity, the mass and the extension of the disk increase with the initial mass (see upper panels of Figs.~\ref{evdisk1} and \ref{evdisk2}). Typically at solar metallicity, the mass of the disk for the 9 $M_{\sun}$ model is a factor of 1600 higher than that of the 1.7 $M_{\sun}$ model, while the extension of the disk is increased by a factor of 2.2 only. Therefore, the disks are expected to be on average denser around more massive stars.
\item For a given initial mass, the extension of the disk increases when the star evolves along the MS as shown in Figs.~\ref{evdisk1} and \ref{evdisk2}. The same is true for the disk diffusion time.
\item For a given initial mass, in general the mass of the disk slightly increases when the metallicity decreases and the extension of the disk slightly decreases. 
For instance, the disk of the 9 $M_{\sun}$ model is 2.8 times more massive at $Z=0.002$ than at $Z=0.014$, while the extension of the disk, in contrast, is 16\% smaller at  $Z=0.002$ than at $Z=0.014$, i.e. some denser disks are expected in low-metallicity regions. 
Overall, Table~\ref{tabla0} shows that the differences due to the metallicity are small compared to those due to the initial mass.
\item The disk diffusion time does not present very strong variations either as a function of the initial mass or as a function of the metallicity.
\end{itemize}

\begin{table*}[t]
\caption{Main characteristics of the decretion disks for models with $\Omega_\text{ini}/\Omega_\text{crit}=0.95$.
}
\label{tabla0}
  \begin{center}
\begin{tabular}{ccccccccccc}
\hline\hline
\rule[0mm]{0mm}{3mm} $Z$ & $M_\text{ini}$ & $\langle\dot{\mathcal{L}}_\text{tot}\rangle$ & $\langle\dot{\mathcal{L}}_{wind}\rangle$ & $\langle\dot{\mathcal{L}}_\text{dw}\rangle$&$\langle\dot{\mathcal{L}}_\text{disk}\rangle$& $\langle\dot{M}_\text{disk}\rangle$ &\multicolumn{2}{c}{$\langle R_\text{crit}\rangle$}&$\langle t_\text{diff}\rangle$&$\langle M_\text{disk}\rangle$ \\
\rule[-1.5mm]{0mm}{3mm}&$M_{\sun}$&g cm$^2$s$^{-2}$& g cm$^2$s$^{-2}$&g cm$^2$s$^{-2}$&g cm$^2$s$^{-2}$&$M_{\sun}\,\text{yr}^{-1}$&$R_{\sun}$&$R_\star$&yr&$M_{\sun}$\\
\hline
\rule[0mm]{0mm}{3mm}$0.014$ & $1.7$ & $2.90\cdot10^{33}$ &          -        &          -        & $2.90\cdot10^{33}$ & $4.04\cdot10^{-13}$ & $2692$ & 738 & $18.2$ & $9.96\cdot10^{-12}$\\
$0.014$ & $2$   & $9.08\cdot10^{33}$ &          -        &          -        & $9.08\cdot10^{33}$ & $1.10\cdot10^{-12}$ & $3416$ & 714 & $23.9$ & $3.10\cdot10^{-11}$\\
$0.014$ & $2.5$ & $2.08\cdot10^{34}$ &          -        &          -        & $2.08\cdot10^{34}$ & $2.26\cdot10^{-12}$ & $3485$ & 652 & $22.1$ & $5.50\cdot10^{-11}$\\
$0.014$ & $3$   & $4.76\cdot10^{34}$ &          -        &          -        & $4.76\cdot10^{35}$ & $4.65\cdot10^{-12}$ & $3470$ & 618 & $20.0$ & $1.08\cdot10^{-10}$\\
$0.014$ & $4$   & $2.72\cdot10^{35}$ &          -        &          -        & $2.72\cdot10^{35}$ & $2.18\cdot10^{-11}$ & $3958$ & 598 & $21.6$ &  $5.17\cdot10^{-10}$\\
$0.014$ & $5$   & $5.67\cdot10^{35}$ &          -        &          -        & $5.67\cdot10^{35}$ & $3.90\cdot10^{-11}$ & $4451$ & 593 & $22.4$ &  $9.25\cdot10^{-10}$\\
$0.014$ & $7$   & $3.34\cdot10^{36}$ & $6.13\cdot10^{34}$ & $3.17\cdot10^{34}$ & $3.28\cdot10^{36}$ & $1.78\cdot10^{-10}$ & $5047$ & 575 & $22.8$ & $4.43\cdot10^{-9}$\\
$0.014$ & $9$   & $1.46\cdot10^{37}$ & $1.27\cdot10^{36}$ & $6.59\cdot10^{35}$ & $1.33\cdot10^{37}$ & $5.92\cdot10^{-10}$ & $6006$ & 569 & $26.2$ & $1.61\cdot10^{-8}$\\
\hline
\rule[0mm]{0mm}{3mm}$0.006$ & $1.7$ & $3.56\cdot10^{33}$ &          -        &          -        & $3.56\cdot10^{33}$ & $5.40\cdot10^{-13}$ & $2401$ & 742 & $15.3$ & $9.38\cdot10^{-12}$\\
$0.006$ & $2$   & $9.67\cdot10^{33}$ &          -        &          -        & $9.67\cdot10^{33}$ & $1.35\cdot10^{-12}$ & $2620$ & 705 & $16.1$ & $2.42\cdot10^{-11}$\\
$0.006$ & $2.5$ & $2.26\cdot10^{34}$ &          -        &          -        & $2.26\cdot10^{34}$ & $2.64\cdot10^{-12}$ & $3006$ & 693 & $17.6$ & $5.12\cdot10^{-11}$\\
$0.006$ & $3$   & $6.58\cdot10^{34}$ &          -        &          -        & $6.58\cdot10^{34}$ & $6.62\cdot10^{-12}$ & $3390$ & 673 & $19.2$ & $1.39\cdot10^{-10}$\\
$0.006$ & $4$   & $7.44\cdot10^{35}$ &          -        &          -        & $7.44\cdot10^{35}$ & $6.08\cdot10^{-11}$ &  $4039$ & 696 & $21.6$ & $1.33\cdot10^{-9}$\\
$0.006$ & $5$   & $4.73\cdot10^{36}$ &          -        &          -        & $4.73\cdot10^{36}$ & $3.32\cdot10^{-10}$ &  $4433$ & 714 & $22.2$ & $7.21\cdot10^{-9}$\\
$0.006$ & $7$   & $4.73\cdot10^{37}$ & $7.09\cdot10^{34}$ & $3.69\cdot10^{33}$ & $4.72\cdot10^{37}$ & $2.74\cdot10^{-9}$ & $4969$ & 693 & $22.3$ & $5.33\cdot10^{-8}$\\
$0.006$ & $9$   & $2.14\cdot10^{37}$ & $1.45\cdot10^{36}$ & $7.48\cdot10^{35}$ & $2.00\cdot10^{37}$ & $9.28\cdot10^{-10}$& $5609$ & 596 & $23.6$ & $2.21\cdot10^{-8}$\\
$0.006$ & $12$  & $1.64\cdot10^{38}$ & $2.17\cdot10^{37}$ & $1.12\cdot10^{37}$ & $1.42\cdot10^{38}$ & $5.28\cdot10^{-9}$ & $6607$ & 546 & $26.2$ &$1.39\cdot10^{-7}$\\
\hline
\rule[0mm]{0mm}{3mm}$0.002$ & $1.7$ & $5.24\cdot10^{33}$ &        -          &          -        & $5.24\cdot10^{33}$ & $8.96\cdot10^{-13}$ & $2005$ & 768 & $11.6$ & $1.19\cdot10^{-11}$\\
$0.002$ & $2$   & $9.32\cdot10^{33}$ &        -          &          -        & $9.32\cdot10^{33}$ & $1.34\cdot10^{-12}$ & $2490$ & 804 & $14.7$ & $2.19\cdot10^{-11}$\\
$0.002$ & $2.5$ & $1.29\cdot10^{35}$ &        -          &          -        & $1.29\cdot10^{35}$ & $1.56\cdot10^{-11}$ & $2975$ & 851 & $17.2$ & $2.68\cdot10^{-10}$\\
$0.002$ & $3$   & $2.91\cdot10^{35}$ &        -          &          -        & $2.91\cdot10^{35}$ & $3.14\cdot10^{-11}$ & $3101$ & 846 & $16.8$ & $4.08\cdot10^{-10}$\\
$0.002$ & $7$   & $5.74\cdot10^{36}$ & $9.48\cdot10^{34}$ & $4.91\cdot10^{34}$ & $5.64\cdot10^{36}$ & $3.36\cdot10^{-10}$ & $4359$ & 703 & $18.3$ & $7.03\cdot10^{-9}$\\
$0.002$ & $9$   & $4.75\cdot10^{37}$ & $1.50\cdot10^{36}$ & $7.72\cdot10^{35}$ & $4.60\cdot10^{37}$ & $2.25\cdot10^{-9}$ & $5029$ & 613 & $20.1$ & $4.53\cdot10^{-8}$\\
$0.002$ & $12$  & $1.01\cdot10^{38}$ & $1.41\cdot10^{37}$ & $7.29\cdot10^{36}$ & $8.69\cdot10^{37}$ & $3.44\cdot10^{-9}$ & $5814$ & 545 & $21.6$ & $7.46\cdot10^{-8}$\\
$0.002$ & $15$  & $1.26\cdot10^{38}$ & $4.90\cdot10^{37}$ & $2.53\cdot10^{37}$ & $7.69\cdot10^{37}$ & $2.57\cdot10^{-9}$ & $6528$ & 533 & $23.1$ & $5.94\cdot10^{-8}$\\
\hline
\hline
\end{tabular}
\end{center}
\tablefoot{Metallicity of the star (column 1), stellar masses (column 2), total angular-momentum loss rate at the critical limit (column 3), angular-momentum loss rate via the stellar wind (column 4), angular-momentum loss via the disk wind considering $R_\text{out} \rightarrow \infty$ (column 5), angular-momentum loss rate via the viscous decretion disk (column 6), mass-loss rate via the decretion disk (column 7), extension of the disk expressed in units of $R_{\sun}$ (column 8) and stellar radius $R_\star$ (column 9), diffusion time considering $\alpha$=1 (column 10), and total mass of the disk (column 11). In all cases an isothermal disk was considered (p=0) and a value of $\alpha$=1.}
\end{table*}

\subsection{Sensitivity to the disk parameters}

It is important to check to what extent the choice of the parameters that describe the thermal structure of the disk and the viscosity parameter affects our results. We use the data of the 7 $M_{\sun}$ model at solar metallicity, two different values for the parameters $T_0/T_\text{eff}$ (0.66 and 0.80), for $p$ (0.0 and 0.1), and for $\alpha$ (0.1 and 1.0). The results are summarised in Table \ref{tablaC2}.

A scatter is produced in the mass, extension, diffusion time, and mass-loss rate of the disk. The highest-to-lowest value ratio varies by a factor of 33 for the mass, 
of 2.3 for the extension, of 20 for the diffusion time and of 1.4 for the mass-loss rate. This illustrates the absolute values obtained in Table \ref{tabla0} are sensitive to the physical ingredients of the disk models. 

The mass and the diffusion time of the disks are the most sensitive quantities, and $\alpha$, the viscosity, is the most critical quantity. A stronger viscosity implies smaller masses for the disks and shorter diffusion times. 

\begin{table}[t]
\caption{Mean values of some disk's characteristics for the $7\,M_{\sun}$ model at solar metallicity, varying $T_{0}/T_\text{eff}$, $p$, and $\alpha$. }
\label{tablaC2}
  \begin{center}
  \scalebox{0.95}{%
\begin{tabular}{ccccccc}
\hline\hline
\rule[0mm]{0mm}{3mm}$T_{0}/T_\text{eff}$&$p$&$\alpha$&$\langle\tau_\text{diff}\rangle$&$\langle R_\text{crit}\rangle$&$\langle\dot{M}_\text{mech}\rangle$&$\langle M_\text{disk}\rangle$\\
\rule[-1.5mm]{0mm}{3mm}&&&$\text{yr}$&$R_{\sun}$&$M_{\sun} \text{yr}^{-1}$&$M_{\sun}$\\
\hline
\rule[0mm]{0mm}{3mm}0.66&0.0&0.1& $228$  & $5047$ & $1.78\cdot10^{-10}$& $4.43\cdot10^{-8}$\\
0.66&0.0&1.0& $22.8$ & $5047$ & $1.78\cdot10^{-10}$& $4.43\cdot10^{-9}$\\
0.66&0.1&0.1& $155$  & $9789$ & $1.28\cdot10^{-10}$& $1.59\cdot10^{-8}$\\
0.66&0.1&1.0& $15.5$ & $9789$ & $1.28\cdot10^{-10}$& $1.59\cdot10^{-9}$\\
0.80&0.0&0.1& $175$  & $4227$ & $1.95\cdot10^{-10}$& $3.71\cdot10^{-8}$\\
0.80&0.0&1.0& $17.5$ & $4227$ & $1.95\cdot10^{-10}$& $3.71\cdot10^{-9}$\\
0.80&0.1&0.1& $117$  & $8038$ & $1.41\cdot10^{-10}$& $1.34\cdot10^{-8}$\\
0.80&0.1&1.0& $11.7$ & $8038$ & $1.41\cdot10^{-10}$& $1.34\cdot10^{-9}$\\
\hline\hline
\end{tabular} }
\end{center}
\tablefoot{Disk parameters (columns 1 to 3), diffusion time (column 4), outer radius (column 5), mechanical mass-loss rate (column 6), and mass of the disk (column 7). The quantities are time-averaged over the critical-rotation phase duration.}
\end{table}

\section{Possible link between very rapid rotation and the classical Be-type star phenomenon \label{secbe}}

\subsection{Is critical rotation a necessary and sufficient condition for Be-type stars?}

There is no doubt that Be stars are rapid rotators, {\it i.e.} with $\omega > 0.70$, however, it is not clear whether Be stars are rotating at the critical limit or not. In the sample studied by \citet{Levenhagen2004a}, eight Be-type stars with a mass estimate between 8 and 10 $M_{\sun}$ show $V \sin i$ values between 160 and 335 km s$^{-1}$, with a mean value of 230 km s$^{-1}$. Corrected for a random orientation of the rotational axis, this value translates into a mean $V_\text{eq}$ around 290 km s$^{-1}$. The critical velocity for our 9 $M_{\sun}$ stellar model at the end of the MS phase (when the most rapidly rotating models reach the critical limit) is 410 km s$^{-1}$. Comparing these two values, we obtain a mean ratio of $V/V_\text{crit}=0.71$ ({\it i.e.} $\omega=0.88$) for the observed stars, which means that Be-type stars may not necessarily be at the critical limit. The same conclusion can be deduced from the work of \citet{Fremat2005a}, who found a mean ratio of $\omega=0.88$ from the fit of photospheric lines of Be stars, accounting for gravity darkening, which also agrees with the result of \citet{Levenhagen2004a}, who did not account for it. Be-shell stars are assumed to be Be stars seen equator-on. Thus, the determination of $V \sin i$ for these objects gives direct information on $V_\text{eq}$. \citet{Rivinius2006a} showed that for the brightest shell stars of different spectral types, the mean fraction of the critical velocity is $81\%$, which corresponds to a mean $\omega$ higher than 0.92. \citet{Meilland2012a} found a mean value of $\omega=0.95$ for a sample of eight Be stars, considering $V \sin i$ and $V_\text{crit}$ from \citet{Fremat2005a} but $i$ measurements derived from interferometry. \citet{Delaa2011a} deduced from interferometric observations of Be stars 48 Per and $\Psi$ Per that both stars rotate at nearly critical rotation. Therefore, the question  to which extent Be stars are really critically rotating stars is not settled yet. Two additional factors have to be accounted for in that discussion: 1) when the surface velocity approaches the critical limit, the method of velocity measurement based on the observed width of the absorption lines might underestimate the values \citep{Townsend2004a}; 2) it may be that Be stars are not always critically rotating, and undergo periods in the course of their evolution during which they rotate below the critical value and periods when they are rotating critically.

\begin{figure}
\centering{
\includegraphics[width=.45\textwidth]{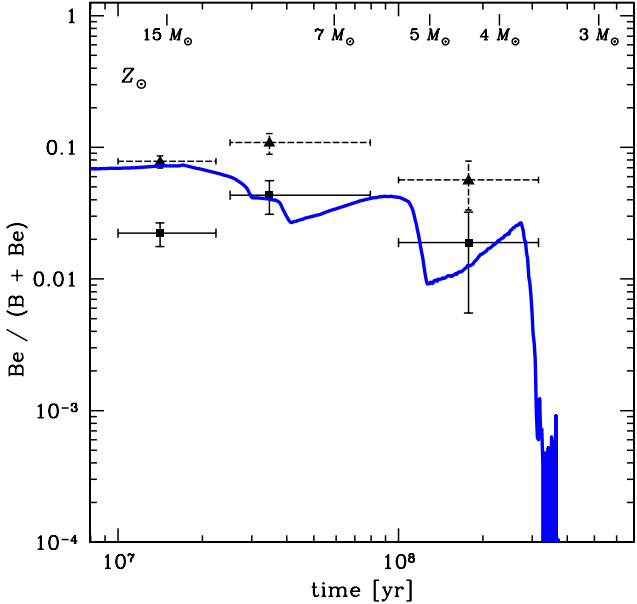}}
\caption{Evolution of the Be/(B+Be) ratio in clusters of various ages at $Z=0.014$ obtained from our models assuming a mass (or $T_\text{eff}$) dependence for the Be phenomenon as proposed by \citet[using $(V_\text{eq}/V_\text{crit})_\text{min}$ values]{Cranmer2005a}. The observational points are the same as in Fig.~\ref{z014H2010}. }
\label{z014H2010Mdep}
\end{figure}

In Fig.~\ref{z014H2010} and \ref{z002H2010}, we show the fraction of Be stars observed in clusters with ages between 10 and 300 Myr at solar metallicity \citep[data from][]{McSwain2005b} and between 10 Myr and 1 Gyr at the SMC metallicity \citep[data extracted from][]{Martayan2010a}, and compare them to the ratio given by our models. Considering only single stars with $\omega > 0.98$, the numbers are much too low to account for the observed fraction of Be-type stars, particularly for young clusters. This last conclusion agrees with the study by \citet{Ekstrom2008b}. Stars with $\omega \sim 0.80$ cannot reproduce the observed trend with age either. Then, in the frame our hypothesis, our present models do not provide a reasonable fit to the observations. 

A different conclusion is obtained when one considers that the lowest velocity that a star has to reach to present the Be phenomenon may vary as a function of the stellar mass and/or the temperature. This was suggested by \citet{Cranmer2005a} and \citet{Huang2010a}, who found that more massive (or hotter) stars become Be-type stars at lower $V/V_\text{crit}$ values than lower-mass (or cooler) stars. For instance, if we consider the lowest rotation rate values $(V_\text{eq}/V_\text{crit})_\text{min}$ in bins of $T_\text{eff}$ proposed by \citet{Cranmer2005a} to obtain a Be star, then the fractions of stars with velocities higher than these values (which appear as Be stars) given by our models reproduce the observed fraction of Be stars quite well (see Fig.~\ref{z014H2010Mdep}). Our models accordingly support the analysis by \citet{Cranmer2005a}.

In addition to the variation of the fraction of Be stars with age, another important feature is the way this fraction varies as a function of metallicity at a given age. It has been established \citep[see also our Fig.~\ref{z014H2010} and \ref{z002H2010}]{Maeder1999a, Wisniewski2006a}
that there is a larger fraction of Be stars at low metallicity. Indeed, 
the observed fraction of Be-type stars in clusters with ages between 10 and 25 Myr is higher by about a factor of 2 at SMC metallicity (Be/(Be+B) fraction of about 30\%) than at solar metallicity (15\%). 
Starting from identical initial velocity distribution, we did not obtain such an enhancement in our models.
Solutions of this problem can come from different directions. For instance, it may be that the initial rotation distribution is shifted to higher velocities at low metallicity. Another possibility could be that since disks are denser at low metallicity (for a given mass and age), they could be more easily detected at low metallicity. 
According to \citet{Silaj2010a}, the minimum density for a disk to produce H$\alpha$ emission is $5\cdot10^{-13}\,\text{g}\ \text{cm}^{-3}$.
Finally, the analysis of \citet{Cranmer2005a}, which sets a minimum velocity for becoming a Be-type star that varies as a function of the effective temperature was performed at solar metallicity. At present we cannot say whether such an analysis would hold at lower metallicities and whether this would help in resolving the above question.

On the basis of our single-star models, we therefore conclude that there are much too few stars that rotate strictly at the critical limit to account for the observed Be-type star population. If the Be phenomenon appears at lower rotation rate, as suggested by many authors, and has a dependence on the mass and/or the $T_\text{eff}$, as proposed by \citet{Cranmer2005a} and \citet{Huang2010a}, the models reproduce the fraction of Be-type stars much better. As a consequence, some other mechanism is required in addition to the centrifugal force to push the matter into a disk.

This conclusion of course depends on the physics included in the models. Here, we did not account for any internal dynamo, which would impose a solid-body rotation during the MS phase and in turn would generate higher surface rotations during the MS phase. Moreover, in addition to the mechanism explored here, where the surface is accelerated by internal processes, some stars may acquire rapid rotation by tidal interactions or through mass exchange in close binaries. Furthermore, as already discussed previously, we started the computation on the ZAMS with a flat $\Omega$ profile, which implies an internal readjustment period at the very beginning of the evolution. As a result, the effective $\omega_\text{ini}$ is lower than the one by which we label our models, and we do not produce critical rotators on the ZAMS.

\subsection{Comparison of the mass-loss rates}

\begin{figure}[h]
\centering{
    \includegraphics[width=.48\textwidth]{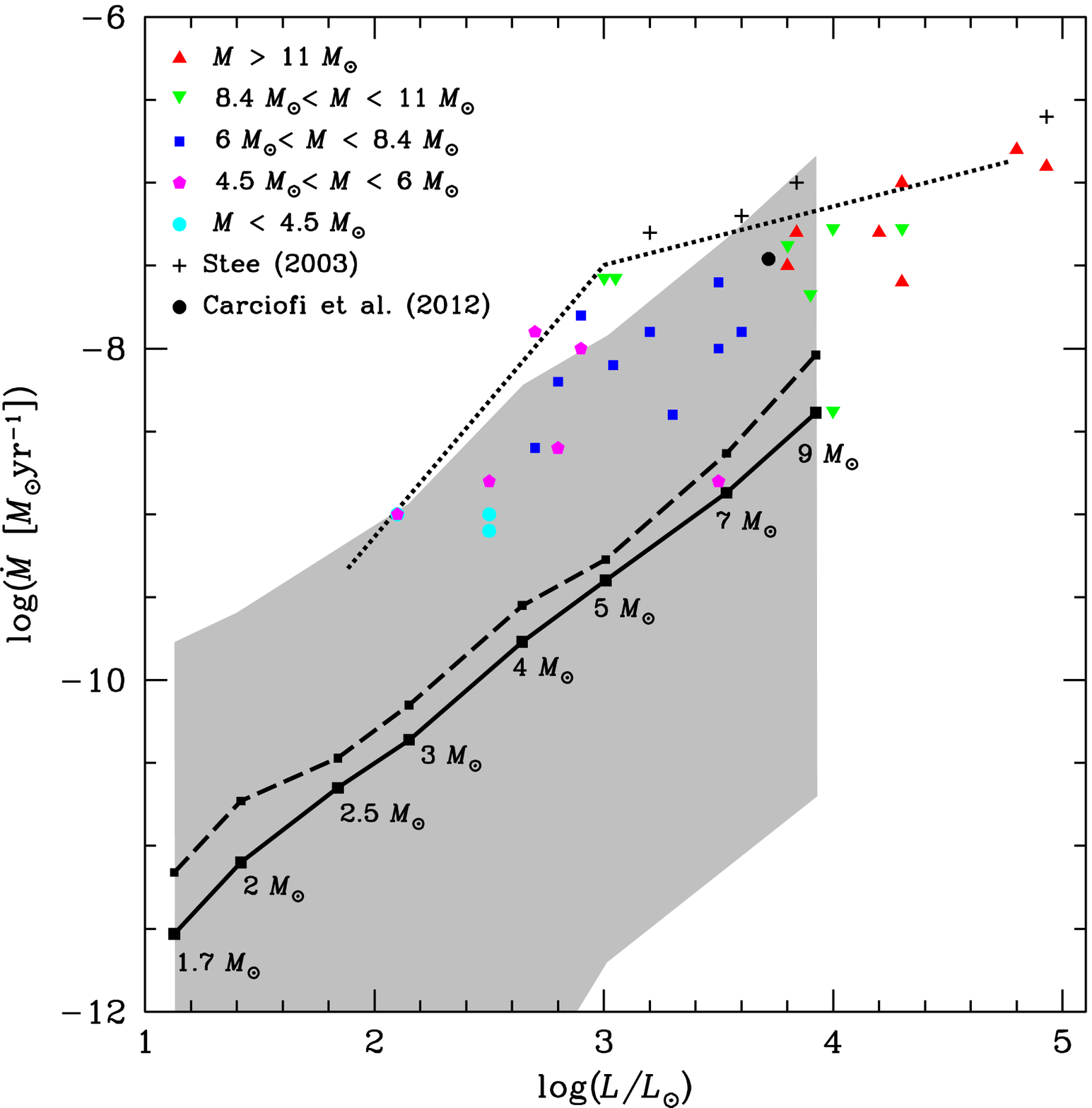}}
    \caption{Mass loss - luminosity diagram. The colour points correspond to data by \citet{Waters1987a}, while the black crosses correspond to data from \citet{Stee2003a} and the black circle to the data from \citet{Carciofi2012a} for 28 CMa. The thin dotted curve traces the upper limit for IR mass-loss rates obtained by \citet{Waters1987a}. The thick black curve corresponds to the mean mechanical mass-loss rates along the critical rotation phase, obtained as described in Section \ref{secmec} and shown in Table \ref{TabBeModels}, while the dashed black line corresponds to the mean value during the time in which the star is effectively losing mass. The grey region corresponds to the range of instantaneous mass-loss rates.}
    \label{Mdot_lum14}
\end{figure}

\citet{Waters1986a} and \citet{Waters1987a} studied the density structure and mass-loss rates of Be stars and interpreted the observed IR excesses in terms of a simple pole-on disk model, which allowed them to estimate upper limits for the IR (equatorial) mass-loss rates between $7\cdot10^{-9}$ and $2\cdot10^{-8}\ M_{\sun}\ \text{yr}^{-1}$, and masses around $1-5\cdot10^{-8}\ M_{\sun}$. The thin dotted line in Fig.~\ref{Mdot_lum14} shows this upper limit for IR mass-loss rates obtained by \citet{Waters1987a}, and the colour points in this figure show some of their data for which reliable stellar-mass determinations are available in the literature. \citet{Stee2003a} also obtained mass-loss rates of some Be stars with SIMECA, a latitude-dependent radiative wind plus dense disk model that produces both spectroscopic and interferometric synthetic data, plotted in crosses in Fig.~ \ref{Mdot_lum14}. \citet{Carciofi2012a} obtained similar values for 28 CMa by modelling the decline rate of the V-band excess. Since the values by \citet{Waters1987a} and \citet{Stee2003a} are strongly model dependent, they cannot be taken as strong constraints of the models. However, it is interesting to compare them with the mass loss rates we obtained in the present work. In Fig.~ \ref{Mdot_lum14} we also plot the mean equatorial mass-loss rates obtained from our critically-rotating models as described in Section \ref{secmec} and presented in Table \ref{TabBeModels} as a solid black line for stars with $\Omega/\Omega_\text{crit}$=0.95 on the ZAMS. In this plot we also represented the averaged mass-loss rates for overcritical episodes (black dashed line). As mentioned in Section \ref{secdisks}, the disk mass-loss rates are lower than these values. 

We see that our mass-loss rate  estimates run below the determinations made by \citet{Waters1987a}. The origin of this shift between the two sets of values is difficult to assess, since both may be model dependent. However, it is remarkable that the models predict an increase of the mechanical loss rate with the luminosity similar to that obtained by  \citet{Waters1987a}.

In this context, it is interesting to note that  {\it per se}, an overcritical velocity is not sufficient to drive a mass loss. When it becomes overcritical, matter will be launched into a Keplerian orbit and therefore will remain gravitationally locked to the star (Jupiter is orbiting around the Sun with a critical velocity!). This matter may even fall back on the star, depending on the forces acting on it. Some additional force (or forces?) have to come into play and provide the momentum needed to drive an effective mass loss from the star and/or an outwards-expanding disk. Here we did not account for any additional forces. If they were present, they would probably increase the mass ``mechanically'' lost by a fast-rotating star, and therefore our present estimate must be considered as a lower limit.

Looking at  Fig.~ \ref{Mdot_lum14}, we can see that the upper values obtained in our models for the instantaneous mass loss rates fall in the range of the observed values. 
However, while the time-averaged mass loss rates are meaningful quantities in the frame of our hypothesis, this is not the case of the instantaneous values, which
are quite model dependent. Therefore, we cannot deduce any firm conclusion from this coincidence.

An interesting point is that stars rotating at or near the critical velocity may or may not have an equatorial disk, but all these stars should present some degrees of anisotropy in the radiative winds. As mentioned in Sect.~\ref{secpoleq}, the radiative mass flux ratio between the equator and the pole reaches 0.15 at most. When mechanical mass loss occurs at the equator, the ratio increases to 60. Some observations have been performed on $\alpha$ Ara (a $9-10\ M_{\sun}$ star estimated to rotate at $V_\text{eq}/V_\text{crit}=0.91$, {\it i.e.} $\omega=0.98-0.99$) and $\alpha$ Eri (Achernar, a $6-8\ M_{\sun}$ star estimated to rotate at $\omega=0.992$). For $\alpha$ Ara, \citet{Chesneau2005a} found a polar flux of $1.7\cdot10^{-9}\ M_{\sun}\ \text{yr}^{-1}\ \text{sr}^{-1}$, with a ratio between the polar and equatorial flux $\Phi_\text{eq}/\Phi_\text{pol}=30$. These numbers have been revised in a second paper \citep{Meilland2007a} to $\Phi_\text{pol}=7\cdot10^{-9}\ M_{\sun}\ \text{yr}^{-1}\ \text{sr}^{-1}$ and $\Phi_\text{eq}/\Phi_\text{pol}=0.03$. For $\alpha$ Eri, \citet{Kanaan2008a} found a polar flux of $4.0\cdot10^{-12}\ M_{\sun}\ \text{yr}^{-1}\ \text{sr}^{-1}$ and a ratio $\Phi_\text{eq}/\Phi_\text{pol}=0.01-1$. The observational range is wide, and for the time being, it seems to be premature to use these kind of observations to put constraints on the models.

\subsection{Comparison of the disk mass, extension, and diffusion time}

In our model, we obtained that the masses of disks increase significantly with the stellar mass. Interestingly enough, \citet{Silaj2010a} modelled H$\alpha$ emission line profiles of Be stars with their non-LTE radiative transfer code BEDISK by fitting disk parameters: their Table 3 shows that disks around B0 type stars are more massive than disks around B8 type stars.

Our models predict extensions for the disk's outer radius of some thousands of solar radii. 
Radio studies provide information on the outer regions of disks surrounding Be stars. 
Using the Very Large Array, \citet{Taylor1990a} carried out a radio survey of 21 Be stars selected for their excess IR emission detected by IRAS. They found that the minimum radius of Be-star disks are of several hundreds to thousands of solar radii. Identical results were obtained by \citet{Clark1998a}, who observed a sample of 13 Be stars of the southern sky with the Australian Telescope Compact Array. The values of a few thousand solar radii we obtain with the different stellar masses and different disk parameters seem to be compatible with the observed quantities. 
The extension of the disk may be dependent on the wavelength used to investigate it. In H$\alpha$, the extensions are much shorter, a few tens of solar radii, since only the portion which is nearby the star and therefore hot and dense enough would be observable, while at longer wavelengths, cooler and shallower parts
of the disk can be seen. Our predictions are valuable for isolated stars. Disk truncation can happen if a close companion star is present.

The diffusion timescales we obtain with our models can be compared to long-term variability cycles observed in Be stars. Some of the observed long-term variations in Be stars, consisting of transitions between normal B and Be phases, or brightness variations \citep[see {\it e.g.}][and references therein]{Porter2003a,Carciofi2011a}, can be understood in terms of the formation and dissipation of a circumstellar disk. They typically last from several months or years to decades. The values of a few of decades we obtain are consistent with these timescales, though different choices of disk parameters can affect this quantity.

\section{Conclusions \label{secconclu}}

With an improved version of the Geneva stellar-evolution code, we have studied the characteristics of initially very rapidly-rotating stars as well as some properties of the equatorial disks that are produced when the surface velocity becomes overcritical.

The main conclusions are:
\begin{itemize}
\item The angular-momentum and mass losses (between $4\cdot10^{-13}$ and $4\cdot10^{-9}\ M_{\sun}\ \text{yr}^{-1}$, depending on the initial mass and metallicity) when stars are at the critical limit are very modest and have no strong impact on the way the star evolves.
\item Nevertheless, signatures of rapid initial rotation on the ZAMS remain present in more evolved stages such as the red-giant or supergiant stages, for instance the higher values for rotation and surface nitrogen enrichment compared to stars that evolved from slowly rotating progenitors.
\item For stars that reach the critical limit, the time-averaged mass-loss rates we obtain are one order of magnitude lower than those currently estimated from observations of Be-type stars. However, among the range of instantaneous mass-loss rates we obtained, the highest values
 agree relatively well with the estimates derived from observations. 
\item When the star is losing mass at the critical velocity, the mass of the disk produced is estimated to range between $9.4\cdot10^{-12}$ and $1.4\cdot10^{-7}\,M_{\sun}$ (from $3\cdot10^{-6}$ to $47\cdot10^{-2}\,M_{\oplus}$), its extension between 2000 and 6500 $R_{\sun}$, and its typical evolution timescale between 10 and 30 yr. Some hints were given on how these values vary as a function of the initial mass, age, and metallicity. With our simple assumptions, even though the mass-loss rates are somewhat lower than those derived from observations, the masses, extensions, and diffusion times of the disk obtained here are consistent with the current estimates of these quantities for observed disks around Be-type stars. 
\item If we assume that there exists a $T_\text{eff}$ dependence for the minimum $\omega$ for the Be phenomenon to occur (as suggested in the literature), the models agree reasonably well with the observed values of Be/(B+Be).
\end{itemize}

Our results were obtained in the frame of the shellular rotation model. If an internal magnetic field would more strongly couple the core to the envelope, and provided no substantial magnetic braking occurs at the surface, the star may reach (at an earlier evolutionary stages) the critical limit more easily. In a forthcoming paper we will explore the consequences of a strong interior coupling.  

\begin{acknowledgements}
The authors thank the referee of this article for his/her constructive comments.
CG acknowledges support from EUFP7-ERC-2012-St Grant 306901.
JK was supported by grant GA \v{C}R 205/08/0003. 
\end{acknowledgements}

\bibliographystyle{aa}
\bibliography{granada_arxiv.bbl}
\end{document}